\PassOptionsToPackage{unicode}{hyperref}
\PassOptionsToPackage{hyphens}{url}
\PassOptionsToPackage{dvipsnames,svgnames,x11names}{xcolor}
\documentclass[
]{article}
\usepackage{xcolor}
\usepackage{amsmath,amssymb}
\usepackage{iftex}
\ifPDFTeX
  \usepackage[T1]{fontenc}
  \usepackage[utf8]{inputenc}
  \usepackage{textcomp} 
  \providecommand{\symbf}[1]{\boldsymbol{#1}}
  \DeclareUnicodeCharacter{2083}{\textsubscript{3}}
\else 
  \usepackage{unicode-math} 
  \defaultfontfeatures{Scale=MatchLowercase}
  \defaultfontfeatures[\rmfamily]{Ligatures=TeX,Scale=1}
\fi
\usepackage{lmodern}
\ifPDFTeX\else
\fi
\IfFileExists{upquote.sty}{\usepackage{upquote}}{}
\IfFileExists{microtype.sty}{
  \usepackage[]{microtype}
  \UseMicrotypeSet[protrusion]{basicmath} 
}{}
\makeatletter
\@ifundefined{KOMAClassName}{
  \IfFileExists{parskip.sty}{%
    \usepackage{parskip}
  }{
    \setlength{\parindent}{0pt}
    \setlength{\parskip}{6pt plus 2pt minus 1pt}}
}{
  \KOMAoptions{parskip=half}}
\makeatother
\makeatletter
\ifx\paragraph\undefined\else
  \let\oldparagraph\paragraph
  \renewcommand{\paragraph}{
    \@ifstar
      \xxxParagraphStar
      \xxxParagraphNoStar
  }
  \newcommand{\xxxParagraphStar}[1]{\oldparagraph*{#1}\mbox{}}
  \newcommand{\xxxParagraphNoStar}[1]{\oldparagraph{#1}\mbox{}}
\fi
\ifx\subparagraph\undefined\else
  \let\oldsubparagraph\subparagraph
  \renewcommand{\subparagraph}{
    \@ifstar
      \xxxSubParagraphStar
      \xxxSubParagraphNoStar
  }
  \newcommand{\xxxSubParagraphStar}[1]{\oldsubparagraph*{#1}\mbox{}}
  \newcommand{\xxxSubParagraphNoStar}[1]{\oldsubparagraph{#1}\mbox{}}
\fi
\makeatother

\usepackage{longtable,booktabs,array}
\usepackage{calc} 
\usepackage{etoolbox}
\makeatletter
\patchcmd\longtable{\par}{\if@noskipsec\mbox{}\fi\par}{}{}
\makeatother
\IfFileExists{footnotehyper.sty}{\usepackage{footnotehyper}}{\usepackage{footnote}}
\makesavenoteenv{longtable}
\usepackage{graphicx}
\makeatletter
\newsavebox\pandoc@box
\newcommand*\pandocbounded[1]{
  \sbox\pandoc@box{#1}%
  \Gscale@div\@tempa{\textheight}{\dimexpr\ht\pandoc@box+\dp\pandoc@box\relax}%
  \Gscale@div\@tempb{\linewidth}{\wd\pandoc@box}%
  \ifdim\@tempb\p@<\@tempa\p@\let\@tempa\@tempb\fi
  \ifdim\@tempa\p@<\p@\scalebox{\@tempa}{\usebox\pandoc@box}%
  \else\usebox{\pandoc@box}%
  \fi%
}
\def\fps@figure{htbp}
\makeatother

\NewDocumentCommand\citeproctext{}{}
\NewDocumentCommand\citeproc{mm}{%
  \begingroup\def\citeproctext{#2}\cite{#1}\endgroup}
\makeatletter
 \let\@cite@ofmt\@firstofone
 \def\@biblabel#1{}
 \def\@cite#1#2{{#1\if@tempswa , #2\fi}}
\makeatother
\newlength{\cslhangindent}
\newlength{\csllabelwidth}
\newenvironment{CSLReferences}[2] 
 {\begin{list}{}{%
  \setlength{\itemindent}{0pt}
  \setlength{\leftmargin}{0pt}
  \setlength{\parsep}{0pt}
  \ifodd #1
   \setlength{\leftmargin}{\cslhangindent}
   \setlength{\itemindent}{-1\cslhangindent}
  \fi
  \setlength{\itemsep}{#2\baselineskip}}}
 {\end{list}}
\usepackage{calc}

\providecommand{\KOMAoption}[2]{}
\usepackage{amssymb}
\usepackage{algorithm}
\usepackage{algpseudocode}
\usepackage{arxiv}
\usepackage{orcidlink}
\usepackage{amsmath}
\usepackage[T1]{fontenc}
\makeatletter
\@ifpackageloaded{caption}{}{\usepackage{caption}}
\AtBeginDocument{%
\ifdefined\contentsname
  \renewcommand*\contentsname{Table of contents}
\else
  \newcommand\contentsname{Table of contents}
\fi
\ifdefined\listfigurename
  \renewcommand*\listfigurename{List of Figures}
\else
  \newcommand\listfigurename{List of Figures}
\fi
\ifdefined\listtablename
  \renewcommand*\listtablename{List of Tables}
\else
  \newcommand\listtablename{List of Tables}
\fi
\ifdefined\figurename
  \renewcommand*\figurename{Figure}
\else
  \newcommand\figurename{Figure}
\fi
\ifdefined\tablename
  \renewcommand*\tablename{Table}
\else
  \newcommand\tablename{Table}
\fi
}
\@ifpackageloaded{float}{}{\usepackage{float}}
\floatstyle{ruled}
\@ifundefined{c@chapter}{\newfloat{codelisting}{h}{lop}}{\newfloat{codelisting}{h}{lop}[chapter]}
\floatname{codelisting}{Listing}

\makeatother
\makeatletter
\@ifpackageloaded{caption}{}{\usepackage{caption}}
\@ifpackageloaded{subcaption}{}{\usepackage{subcaption}}
\makeatother
\usepackage{bookmark}
\IfFileExists{xurl.sty}{\usepackage{xurl}}{} 
\hypersetup{
  pdftitle={The BiP-PRISM algorithm for fast and scalable core-loss STEM-EELS simulations},
  pdfauthor={Philipp Pelz},
  pdfkeywords={STEM-EELS, multislice, PRISM, scattering
matrix, simulation, core-loss},
  colorlinks=true,
  linkcolor={blue},
  filecolor={Maroon},
  citecolor={Blue},
  urlcolor={Blue},
  pdfcreator={LaTeX via pandoc}}

\newcommand{\runninghead}{A Preprint }
\title{The BiP-PRISM algorithm for fast and scalable core-loss STEM-EELS
simulations}
\author{\textbf{Philipp
Pelz}~\orcidlink{0000-0002-8009-4515}\\Friedrich-Alexander-Universität
Erlangen-Nürnberg\\Department of Materials Science and
Engineering\\Erlangen\\\href{mailto:philipp.pelz@fau.de}{philipp.pelz@fau.de}}
\date{}
\begin{document}
\maketitle
\begin{abstract}
Quantitative interpretation of atomic-resolution STEM-EELS requires
dynamical simulation of the electron probe before and after core-loss
transitions, which is computationally expensive.

While the PRISM algorithm accelerates this by reusing scattering
matrices, we introduce beam partitioning for both the probe-forming
(\(\mathcal{S}_1\)) and detector-propagating (\(\mathcal{S}_2\)) PRISM
matrices to further reduce computational and memory costs. Each matrix
is calculated on a sparse set of parent beams and reconstructed via
natural-neighbor interpolation locally at the ionized atom.

A locality result demonstrates that the total error is governed entirely
by this on-atom reconstruction error. The resulting BiP-PRISM algorithm
removes per-scan exit wave propagation and significantly reduces memory
requirements, enabling full-resolution elemental mapping, 4D cubes, and
momentum-resolved qEELS on consumer-grade GPUs.

We characterize the approximation's validity regime and demonstrate the
simulation of a multimodal five-edge oxide-interface map and an FePt
nanoparticle Fe-L map at 5x memory reduction, showing that the algorithm
achieves high accuracy with significantly lower computational demands.
\end{abstract}
{\bfseries \emph Keywords}
\def\sep{\textbullet\ }
STEM-EELS \sep multislice \sep PRISM \sep scattering
matrix \sep simulation \sep 
core-loss

\section{Introduction}\label{sec-intro}

Transmission electron microscopy is a powerful tool for studying
materials at the atomic scale. Combining scanning transmission electron
microscopy (STEM) with electron energy loss spectroscopy (EELS) enables
atomic-resolution elemental mapping. However, strong dynamical (multiple
elastic) scattering of the focused electron probe prevents direct map
interpretation. Quantitative interpretation requires full
quantum-mechanical simulation of the electron probe before and after the
core-loss transition (\citeproc{ref-dwyer2005}{Dwyer 2005};
\citeproc{ref-dwyer2008}{Dwyer, Findlay, and Allen 2008};
\citeproc{ref-brown2019}{Brown, Ciston, and Ophus 2019}). Several
simulation packages perform these dynamical calculations, including the
C++/Fortran-based muSTEM (\citeproc{ref-allen2015}{Allen, D'Alfonso, and
Findlay 2015}) and the Python-based abTEM
(\citeproc{ref-madsen2021}{Madsen and Susi 2021}) and py\_multislice
(pyms) (\citeproc{ref-brown2020}{Brown et al. 2020}).

The conventional transition-potential multislice algorithm repeats the
elastic multislice propagation for each probe position, propagating an
independent inelastic wave to the exit surface for every combination of
probe position, transition channel, and ionized atom. This method scales
poorly for the large, heterogeneous fields of view typically studied
experimentally, requiring days or weeks for modest simulation cells
(\citeproc{ref-brown2019}{Brown, Ciston, and Ophus 2019}).

To reduce these computational costs, Brown, Ciston, and Ophus
(\citeproc{ref-brown2019}{2019}) extended the plane-wave
reciprocal-space interpolated scattering matrix (PRISM) algorithm
(\citeproc{ref-ophus2017}{Ophus 2017}) to STEM-EELS simulations. This
algorithm precomputes a probe-forming scattering matrix
\(\mathcal{S}_1\) and reuses it across all scan positions, achieving
near-linear scaling without additional error. It also allows beam
subsampling to trade a controlled amount of accuracy for speed. Other
parallelization strategies for inelastic scattering include the
phase-scrambling approach introduced by Mendis
(\citeproc{ref-mendis2023}{2023}).

For elastic STEM, Pelz, Rakowski, et al. (\citeproc{ref-pelz2022}{2021})
introduced a beam partitioning algorithm that projects the scattering
matrix onto a sparse set of parent beams and reconstructs the full
aperture using natural-neighbor interpolation. This reduces both the
required multislice propagations and the resident memory usage.

In this work, we partition both scattering matrices required for
PRISM-EELS simulations: the probe-forming matrix \(\mathcal{S}_1\) and
the detector-side matrix \(\mathcal{S}_2\). We present the
partitioned-PRISM-EELS algorithms (Section~\ref{sec-algorithms}),
highlighting a double-partitioned elemental-map variant that compresses
both \(\mathcal{S}_1\) and the adjoint detector matrix \(\mathcal{S}_2\)
(Section~\ref{sec-double}) to yield the greatest speed and memory gains.
We also present a locality theorem showing that because the inelastic
coupling is localized to a small window around the ionized atom, the
total simulation error is governed entirely by the on-atom
reconstruction error (Section~\ref{sec-locality}). Furthermore, we unify
the treatment of different output modes (elemental maps,
momentum-resolved qEELS, and 4D cubes) as reductions over the detector
beams (Section~\ref{sec-outputs}). Finally, we characterize the
approximation's validity regime across edge depth, thickness, and
defocus, and evaluate its performance (Section~\ref{sec-experiments}).
We demonstrate a \(\sim\!5\times\) peak-memory reduction on a
23,000-atom FePt nanoparticle and simulate a five-edge multimodal
oxide-interface map.

\section{Methods}\label{sec-methods}

\subsection{STEM-EELS forward model}\label{sec-theory}

We establish notation and describe the energy-filtered forward model
computed by the algorithms in Section~\ref{sec-algorithms}. We follow
the scattering-matrix notation of Pelz, Rakowski, et al.
(\citeproc{ref-pelz2022}{2021}) throughout, scan position
\(\symbf{\rho}\), beams \(\mathbf h\), probe-forming aperture
\(\Psi(\mathbf h)\), scattering matrix \(\mathcal S\), parent beams, and
beamlet basis \(\widehat\psi_p\), extended to the inelastic (EELS) case.
A glossary of symbols is given in Table~\ref{tbl-notation}
(Section~\ref{sec-notation}).

\subsubsection{Grid, probe and specimen}\label{grid-probe-and-specimen}

The specimen is sampled on a real-space grid of \(N_y \times N_x\)
pixels spanning a field of view \((L_y, L_x)\), and sliced into \(N_Z\)
slices of thickness \(\Delta z\) along the beam direction \(z\). Here,
\(\mathbf r\) represents a real-space pixel coordinate, \(\mathbf q\)
denotes a spatial frequency, and \(\mathbf h\) is the integer index of a
beam on the corner-origin discrete Fourier grid.

The probe-forming aperture is a complex function \(\Psi(\mathbf h)\) in
reciprocal space (the lens transfer function inside the aperture, zero
outside). Its support defines the set of \(B\) aperture \emph{beams}
\(\{\mathbf h_b\}_{b=1}^{B} = \{\mathbf h : \Psi(\mathbf h) \neq 0\}\)
with coefficients \(c_b \equiv \Psi(\mathbf h_b)\). A probe centred at
scan position \(\symbf{\rho}\) (in pixels) is the inverse transform of
the aperture with a linear phase ramp,

\begin{equation}\phantomsection\label{eq-probe}{
\psi_0(\mathbf r, \symbf{\rho})
  = \sum_{b=1}^{B} c_b \,
    e^{-2\pi i\,\mathbf h_b\cdot \symbf{\rho} / \mathbf N}\,
    e^{\,2\pi i\,\mathbf h_b\cdot \mathbf r / \mathbf N},
}\end{equation}

where
\(\mathbf h\cdot\symbf{\rho}/\mathbf N \equiv h_y \rho_y/N_y + h_x \rho_x/N_x\).
PRISM exploits the identity in Equation~\ref{eq-probe}: the probe is a
fixed linear combination of plane waves where the scan position
\(\symbf{\rho}\) enters only as a per-beam phase ramp.

The interaction with slice \(j\) is the transmission function
\(T_j(\mathbf r) = \exp\!\big(i\sigma V_j(\mathbf r)\big)\), with
\(\sigma\) the interaction constant and \(V_j\) the projected potential
of the slice under a single frozen-phonon configuration (the final
intensity is averaged over multiple configurations). Free-space
propagation between slices is the Fresnel operator \(\mathcal P\) with
transfer function

\begin{equation}\phantomsection\label{eq-propagator}{
P(\mathbf q) = \exp\!\big(-i\pi\lambda\,\Delta z\,|\mathbf q|^2\big),
\qquad
\mathcal P\,\psi = \mathcal F^{-1}\!\big[\,P\cdot \mathcal F\psi\,\big],
}\end{equation}

where \(\lambda\) is the electron wavelength and \(\mathcal F\) the 2D
DFT. One multislice step through slice \(j\) is transmit-then-propagate,
\(\psi \mapsto \mathcal P\,(T_j\,\psi)\).

\subsubsection{Core-loss transition and the energy-filtered
cube}\label{core-loss-transition-and-the-energy-filtered-cube}

Our model adopts the transition-potential formulation of inner-shell
ionization (\citeproc{ref-dwyer2005}{Dwyer 2005}) under the
single-inelastic-scattering approximation. For an ionizable atom of the
target element at fractional in-plane position \(\symbf{\tau}\) (lying
in slice \(i(\symbf{\tau})\)), the inelastic event with final-state
channel \(n\) converts the elastic wave \(\psi\) arriving at that plane
into an inelastically-scattered wave

\begin{equation}\phantomsection\label{eq-inelastic}{
\psi_n(\mathbf r) = H_{n0}(\mathbf r - \symbf{\tau})\,\psi(\mathbf r),
}\end{equation}

where \(H_{n0}\) is the transition potential, the matrix element
coupling the bound initial state \(0\) to continuum final state \(n\)
for the chosen edge and energy window. The precomputed transition
potentials \(\{H_{n0}\}_{n=1}^{N_{\mathrm{ch}}}\) are sharply localized
about the ionized atom (sub-angstrom support), driving the locality
result in Section~\ref{sec-locality}.

We propagate each inelastic wave to the exit surface via multislice and
Fourier-transform the result to the detector. Writing
\(\mathcal M_{i\to \mathrm{exit}}\) for the multislice exit operator
from slice \(i\) (transmit-and-propagate through slices
\(i,\dots,N_Z-1\)), the energy-filtered diffraction intensity recorded
at scan position \(\symbf{\rho}\) is

\begin{equation}\phantomsection\label{eq-cube}{
I(\mathbf q, \symbf{\rho})
  = \sum_{\symbf{\tau}}\;\sum_{n=1}^{N_{\mathrm{ch}}}
    \Big|\,\mathcal F\,\mathcal M_{i(\symbf{\tau})\to\mathrm{exit}}
      \big[\,H_{n0}(\cdot-\symbf{\tau})\,\psi(\cdot,\symbf{\rho})\,\big](\mathbf q)\,\Big|^2,
}\end{equation}

summed incoherently over all target-element sites \(\symbf{\tau}\) in
the field of view and over all transition channels \(n\) (and, finally,
averaged over frozen-phonon configurations). Stacking
Equation~\ref{eq-cube} over the \(P\) scan positions \(\symbf{\rho}\)
yields the energy-filtered 4D-STEM cube, two scan dimensions \(\times\)
two detector dimensions, for the chosen ionization edge.

\subsubsection{Output modes}\label{sec-outputs-theory}

Because experiments rarely record the full 4D cube, we define two common
reductions of Equation~\ref{eq-cube} over the detector coordinate
\(\mathbf q\):

\begin{itemize}
\item
  Elemental map \(I(\symbf{\rho})\), the energy-filtered intensity
  collected over a detector of semi-angle \(\theta_{\text{det}}\), the
  standard STEM-EELS image:

  \begin{equation}\phantomsection\label{eq-map}{
  I(\symbf{\rho})=\!\!\sum_{|\mathbf q|\le\theta_{\text{det}}/\lambda}\!\! I(\mathbf q,\symbf{\rho}).
  }\end{equation}
\item
  qEELS (momentum-resolved) \(I(q_\parallel,\symbf{\rho})\), one
  diffraction axis resolved, the perpendicular axis projected away:
  \(I(q_\parallel,\symbf{\rho})=\sum_{q_\perp} I(\mathbf q,\symbf{\rho})\).
\item
  4D cube \(I(\mathbf q,\symbf{\rho})\), no reduction
  (Equation~\ref{eq-cube}).
\end{itemize}

These are the common output of the algorithms below; they differ only in
how the elastic probe \(\psi(\cdot,\symbf{\rho})\) at the ionization
plane is formed and, for the map and qEELS, how the exit propagation to
the detector is represented. As Section~\ref{sec-algorithms} shows, the
finite detector of Equation~\ref{eq-map} is exactly what lets the exit
side collapse onto a second, reusable scattering matrix.

Crucially, the scan position \(\symbf{\rho}\) in Equation~\ref{eq-cube}
enters only via the elastic wave \(\psi(\cdot,\symbf{\rho})\) at the
ionization plane. Everything downstream involves exact unitary
propagation and is identical across all algorithms. Accelerating the
simulation therefore reduces to accelerating the computation of
\(\psi(\cdot,\symbf{\rho})\) for many \(\symbf{\rho}\), and any
approximation made there is confined by the locality of \(H_{n0}\).

\subsection{Algorithms}\label{sec-algorithms}

We present the algorithms in order of increasing computational reuse,
building toward our primary method. Conventional multislice
(Section~\ref{sec-conventional}) reuses no intermediate results. PRISM
(Section~\ref{sec-prism}) reuses the probe-forming matrix
\(\mathcal S_1\) across scan positions, and beam partitioning compresses
this matrix onto parent beams (Section~\ref{sec-partitioned}). For
detector-integrated elemental maps, a second scattering matrix
\(\mathcal S_2\) represents the exit side (Section~\ref{sec-dual});
partitioning both matrices (Section~\ref{sec-double}) yields the
greatest speed and memory gains. The energy-filtered 4D cube and
momentum-resolved qEELS outputs are computed as reductions over detector
beams (Section~\ref{sec-outputs}). Throughout, the algorithms maintain
the underlying physics of Equation~\ref{eq-cube}--Equation~\ref{eq-map},
differing only in how the probe and the exit propagation are represented
and reused.

\begin{figure}[H]

\centering{

\includegraphics[width=1\linewidth,height=\textheight,keepaspectratio]{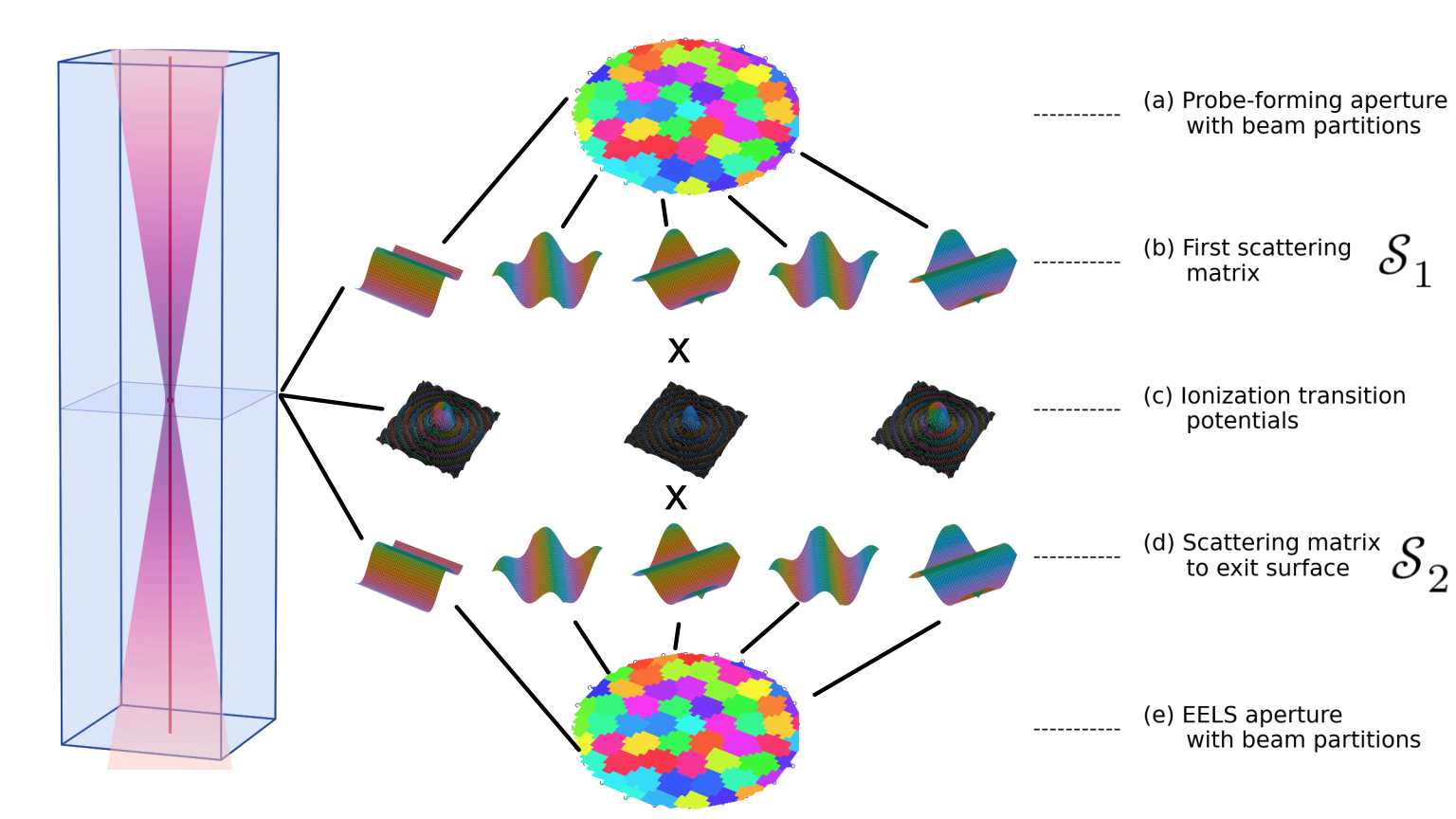}

}

\caption{\label{fig-schematic}Double-partitioned PRISM-EELS at a glance.
Bi-partitioned S-Matrix pipeline for detector-integrated elemental
mapping (left: convergent probe traversing a thick specimen; the
ionization plane lies at mid-depth). (a) Entrance aperture with probe
beams grouped into a sparse hex-ring parent set (\(B_{p,1}\ll B\)); the
full probe is recovered on the inelastic window by natural-neighbor
weights. (b) Probe-forming scattering matrix \(\mathcal S_1\): each
parent column is one precomputed multislice propagation to the
ionization depth, shared across all scan positions. (c) Core-loss
transition potentials \(H_{n0}\) (Equation~\ref{eq-inelastic}),
localized at the ionized atom and applied to the reconstructed probe.
(d) Exit scattering matrix \(\mathcal S_2\): parent detector columns
obtained by adjoint multislice back-propagation from the exit surface;
partitioning \(\mathcal S_2\) replaces the per-scan exit propagation of
conventional multislice. (e) Partitioned EELS collection aperture
(\(B_{p,2}\) parents); summation over final-state channels and parent
detector beams yields the scan-pixel intensity.}

\end{figure}

Figure~\ref{fig-schematic} illustrates the proposed double-partitioned
pipeline end-to-end. Table~\ref{tbl-algorithms} lists the recommended
algorithm for each output mode: elemental maps and qEELS use the
double-partitioned Algorithm 5, whereas the 4D cube uses the
\(\mathcal{S}_1\)-partitioned Algorithm 3. Because the 4D cube does not
sum over detector coordinates, \(\mathcal{S}_2\) cannot be compressed,
leaving only the probe-forming matrix \(\mathcal{S}_1\) partitioned.
Algorithm 1 (conventional multislice) serves as the ground truth, while
the unpartitioned Algorithms 2 and 4 serve as the exact references
(oracles) to validate the partitioned methods.

\begin{longtable}[]{@{}
  >{\raggedright\arraybackslash}p{(\linewidth - 6\tabcolsep) * \real{0.2500}}
  >{\raggedright\arraybackslash}p{(\linewidth - 6\tabcolsep) * \real{0.2500}}
  >{\raggedright\arraybackslash}p{(\linewidth - 6\tabcolsep) * \real{0.2500}}
  >{\raggedright\arraybackslash}p{(\linewidth - 6\tabcolsep) * \real{0.2500}}@{}}
\caption{Which algorithm for which output. The map and qEELS partition
both scattering matrices (Algorithm 5); the 4D cube partitions only
\(\mathcal S_1\) (Algorithm 3), because retaining every detector beam
precludes reducing \(\mathcal S_2\). Algorithm 1 is the ground-truth
reference; Algorithms 2 and 4 are the exact, unpartitioned
references.}\label{tbl-algorithms}\tabularnewline
\toprule\noalign{}
\begin{minipage}[b]{\linewidth}\raggedright
Desired output
\end{minipage} & \begin{minipage}[b]{\linewidth}\raggedright
Recommended algorithm
\end{minipage} & \begin{minipage}[b]{\linewidth}\raggedright
Partitioned
\end{minipage} & \begin{minipage}[b]{\linewidth}\raggedright
Exact oracle
\end{minipage} \\
\midrule\noalign{}
\endfirsthead
\toprule\noalign{}
\begin{minipage}[b]{\linewidth}\raggedright
Desired output
\end{minipage} & \begin{minipage}[b]{\linewidth}\raggedright
Recommended algorithm
\end{minipage} & \begin{minipage}[b]{\linewidth}\raggedright
Partitioned
\end{minipage} & \begin{minipage}[b]{\linewidth}\raggedright
Exact oracle
\end{minipage} \\
\midrule\noalign{}
\endhead
\bottomrule\noalign{}
\endlastfoot
Elemental map & Alg. 5, double-partitioned (Section~\ref{sec-double}) &
\(\mathcal S_1\) and \(\mathcal S_2\) & Alg. 4
(Section~\ref{sec-dual}) \\
Momentum-resolved qEELS & Alg. 5, double-partitioned
(Section~\ref{sec-double}) & \(\mathcal S_1\) and \(\mathcal S_2\) &
Alg. 4 (Section~\ref{sec-dual}) \\
Energy-filtered 4D cube & Alg. 3, partitioned \(\mathcal S_1\)
(Section~\ref{sec-partitioned}) & \(\mathcal S_1\) only
(\(\mathcal S_2\) not reducible) & Alg. 2 (Section~\ref{sec-prism}) \\
\end{longtable}

\subsubsection{Conventional transition-potential
multislice}\label{sec-conventional}

The baseline (\citeproc{ref-dwyer2005}{Dwyer 2005};
\citeproc{ref-brown2019}{Brown, Ciston, and Ophus 2019}) forms the probe
Equation~\ref{eq-probe} explicitly at every scan position and propagates
it through the specimen by multislice. At each slice containing a target
atom, every transition channel spawns an inelastic wave
Equation~\ref{eq-inelastic}, which is propagated independently to the
exit surface and accumulated as diffraction intensity (Algorithm 1).
Because both elastic and exit propagations are repeated for every probe
position, the entire simulation scales with the number of scan positions
\(P\). This is the computational bottleneck that Brown, Ciston, and
Ophus (\citeproc{ref-brown2019}{2019}) sought to eliminate. It is exact
within the frozen-phonon and single-inelastic-scattering approximations
and serves as our ground truth.

\begin{algorithm}[ht]
\caption{Conventional transition-potential multislice STEM-EELS (ground truth)}
\label{alg-conventional}
\begin{algorithmic}[1]
\Require aperture $A$; transmissions $\{T_j\}_{j=0}^{N_Z-1}$; propagator $\mathcal P$;
         transition potentials $\{H_n\}$; sites $\{\symbf{\tau}\}$; $P$ scans $\{\symbf{\rho}\}$
\Ensure energy-filtered cube $I(\mathbf q,\symbf{\rho})$
\State $I \gets 0$
\For{each scan position $\symbf{\rho}$}
  \State $\psi \gets \psi_0(\cdot,\symbf{\rho})$ \Comment{form probe, \eqref{eq-probe}}
  \For{slice $i = 0,\dots,N_Z-1$}
    \For{each site $\symbf{\tau}$ in slice $i$, each channel $n$}
      \State $\psi_n \gets H_n(\cdot-\symbf{\tau})\cdot \psi$ \Comment{inelastic wave, \eqref{eq-inelastic}}
      \State $\psi_n \gets \mathcal M_{i\to\mathrm{exit}}(\psi_n)$ \Comment{multislice to exit}
      \State $I(\cdot,\symbf{\rho}) \mathrel{+}= |\mathcal F\psi_n|^2$
    \EndFor
    \State $\psi \gets \mathcal P\,(T_i\,\psi)$ \Comment{advance elastic wave one slice}
  \EndFor
\EndFor
\end{algorithmic}
\end{algorithm}

\subsubsection{PRISM-EELS: the probe-forming scattering
matrix}\label{sec-prism}

Exact PRISM-EELS (Algorithm 2) reuses a single probe-forming scattering
matrix across all scan positions; it is the unpartitioned baseline that
partitioning compresses, and the exact reference against which we
validate it. PRISM (\citeproc{ref-ophus2017}{Ophus 2017};
\citeproc{ref-brown2019}{Brown, Ciston, and Ophus 2019}) exploits the
linearity of multislice and the structure of Equation~\ref{eq-probe}.
Let \(\mathcal S\) be the scattering matrix where the \(b\)-th row is
the multislice propagation of a unit plane wave from aperture beam
\(\mathbf h_b\). Advancing every row by the same slices,

\begin{equation}\phantomsection\label{eq-smatrix-advance}{
\mathcal S_b \;\xleftarrow{\ \text{transmit + propagate}\ }\;
\mathcal P\,(T_j\,\mathcal S_b),
}\end{equation}

and stopping at slice \(i\) yields, for every beam, the elastic field
that a plane wave \(\mathbf h_b\) would have at that depth. Because the
probe Equation~\ref{eq-probe} is a fixed linear combination of those
same plane waves, the propagated elastic probe at plane \(i\) for
\emph{any} scan position is the same linear combination of the advanced
rows,

\begin{equation}\phantomsection\label{eq-prism-probe}{
\psi(\mathbf r,\symbf{\rho})
  = \sum_{b=1}^{B} c_b\,
    e^{-2\pi i\,\mathbf h_b\cdot\symbf{\rho}/\mathbf N}\;
    \mathcal S_b(\mathbf r).
}\end{equation}

Built and advanced once, the matrix is reused across all \(P\) scan
positions (Algorithm 2); only the inexpensive phase-ramp recombination
in Equation~\ref{eq-prism-probe} is computed per position. This enables
PRISM's near-linear scaling by eliminating the per-probe elastic
multislice propagation of Section~\ref{sec-conventional}.

Beam subsampling (interpolation factor). Following Ophus
(\citeproc{ref-ophus2017}{2017}), retaining only every \(f_y\)-th and
\(f_x\)-th beam reduces the number of propagated rows, which decreases
the build cost and memory by a factor of \(f_y f_x\). The synthesized
probe (Equation~\ref{eq-prism-probe}) becomes periodic with period
\((N_y/f_y, N_x/f_x)\), matching the exact probe inside this window.
Since the localized transition potential (Equation~\ref{eq-inelastic})
samples the probe only at the ionized atom, only the probe value at that
atom must lie within the alias-free window; setting \(f=1\) retains all
beams and is exact. (To keep the synthesized intensity on the same
scale, the retained coefficients are rescaled by \(f_y f_x\), the DFT
decimation factor.)

Whether the exit-side computation can be reduced depends on the desired
output mode. For an energy-filtered 4D cube, the full exit diffraction
pattern is required at each scan position. Consequently, the
post-ionization propagation cannot be collapsed into a second scattering
matrix because that would require representing every output beam.
Instead, we propagate each inelastic wave to the exit surface using
conventional multislice (as in Section~\ref{sec-conventional}). PRISM
therefore accelerates only the probe-forming stage, and the per-scan
exit loop establishes a performance floor on the overall speedup
(Section~\ref{sec-complexity}). For detector-integrated elemental
maps---the most common STEM-EELS output---the signal is collected over a
finite aperture, allowing the exit-side propagation to collapse onto a
second scattering matrix, \(\mathcal{S}_2\). This matrix can also be
partitioned to maximize performance gains, as described in
Section~\ref{sec-dual} and Section~\ref{sec-double}. The 4D cube and
momentum-resolved qEELS outputs are recovered as reductions over
detector beams in Section~\ref{sec-outputs}.

\begin{algorithm}[ht]
\caption{PRISM-EELS with the exact per-pixel scattering matrix $\mathcal S_1$}
\label{alg-prism}
\begin{algorithmic}[1]
\Require aperture $A$; transmissions $\{T_j\}$; propagator $\mathcal P$;
         interpolation factor $(f_y,f_x)$; $\{H_n\}$, sites, scans
\Statex Build $\mathcal S_1$ (once, shared across all scans):
\State beams $\gets \{\mathbf h : \Psi(\mathbf h)\neq 0\}$;\ \ keep $\mathbf h$ with $h_y\equiv 0\,(f_y),\,h_x\equiv 0\,(f_x)$
\State $\mathcal S_b \gets \mathcal F^{-1}[\delta_{\mathbf h_b}]$ for each retained beam \Comment{unit plane waves}
\Statex Per ionization plane $i$ (in increasing depth):
\State \textsc{AdvanceTo}($i$): apply \eqref{eq-smatrix-advance} for the new slices
\State probes $\gets \big\{\sum_b c_b\,e^{-2\pi i\,\mathbf h_b\cdot\symbf{\rho}/\mathbf N}\,\mathcal S_b\big\}_{\symbf{\rho}}$ \Comment{\eqref{eq-prism-probe}}
\State accumulate exit intensities as in Alg.~\ref{alg-conventional}, lines 5--8
\end{algorithmic}
\end{algorithm}

\subsubsection{Partitioned PRISM-EELS}\label{sec-partitioned}

Partitioned PRISM compresses the probe-forming matrix: instead of one
propagated row per aperture pixel (\(B\) rows), it propagates only
\(B_p \ll B\) parent rows and reconstructs the full-aperture probe by
interpolation (\citeproc{ref-pelz2022}{Pelz, Rakowski, et al. 2021};
\citeproc{ref-pelz2021}{Pelz, Brown, et al. 2021}). We develop the
compression on \(\mathcal S_1\) here; Section~\ref{sec-double} applies
the identical machinery to the exit matrix \(\mathcal S_2\), and the two
together give the headline speed and memory gains.

On its own, with the transition-scattered wave propagated to the exit
surface per scan position, exactly as in Section~\ref{sec-conventional},
this \(\mathcal S_1\)-only partitioning is the algorithm for the
energy-filtered 4D cube (Algorithm 3): because the cube retains every
exit/detector beam, no detector sum is taken and \(\mathcal S_2\) cannot
be introduced. The detector-integrated elemental map and qEELS instead
collapse the exit side onto a second matrix \(\mathcal S_2\)
(Section~\ref{sec-dual}) and partition it as well
(Section~\ref{sec-double}).

Parent selection. The parents are a hex-ring subsample of the aperture:
the DC beam plus \(n_{\text{radial}}\) concentric rings, with
\(n_{\text{angular}}(1+i)\) angular samples on ring \(i\), each snapped
to the nearest aperture beam and deduplicated (Algorithm 3, build step).
This yields \(B_p\) parents \(\{\mathbf h_p\}\) approximately uniformly
covering the disc.

Natural-neighbor weights. Each aperture beam \(\mathbf h_b\) is written
as a local convex combination of the parents using Sibson
natural-neighbor interpolation (\citeproc{ref-sibson1981}{Sibson 1981})
on the signed-frequency coordinates,

\begin{equation}\phantomsection\label{eq-nnw}{
\Psi(\mathbf h_b)\,(\cdots)\;\approx\;\sum_{p=1}^{B_p} w_{p,b}\,(\cdots),
\qquad w_{p,b}\ge 0,\ \ \sum_p w_{p,b}=1,
}\end{equation}

with \(w_{p,b}\) nonzero only for parents surrounding \(\mathbf h_b\).
Following Pelz, Rakowski, et al. (\citeproc{ref-pelz2022}{2021}), the
weight matrix \(\mathbf w \in \mathbb R^{B_p \times B}\) depends solely
on the aperture geometry. Because it is independent of probe defocus,
specimen transmission, and thickness, it is computed once per aperture
and cached for any simulation using that aperture.

De-tilt + beamlet basis. Combining Equation~\ref{eq-prism-probe} with
Equation~\ref{eq-nnw}, the synthesized probe factorizes into a sum over
parents of a de-tilted parent column times the beamlet basis
\(\widehat\psi_p\) of Pelz, Rakowski, et al.
(\citeproc{ref-pelz2022}{2021}):

\begin{equation}\phantomsection\label{eq-part-probe}{
\psi(\mathbf r,\symbf{\rho})
   = \sum_{p=1}^{B_p}\, \widehat\psi_p(\mathbf r,\symbf{\rho})\;
        \mathcal S^{\mathrm{dt}}_p(\mathbf r)
}\end{equation}

\begin{equation}\phantomsection\label{eq-part-basis}{
\mathcal S^{\mathrm{dt}}_p(\mathbf r)
  = \mathcal S_p(\mathbf r)\,e^{-2\pi i\,\mathbf h_p\cdot\mathbf r/\mathbf N},
\qquad
\widehat\psi_p(\mathbf r,\symbf{\rho})
  = \sum_{b=1}^{B} w_{p,b}\,\Psi(\mathbf h_b)\,
     e^{\,2\pi i\,\mathbf h_b\cdot(\mathbf r-\symbf{\rho})/\mathbf N}
}\end{equation}

(\citeproc{ref-pelz2022}{Pelz, Rakowski, et al. 2021}, Eq. 17), computed
as the inverse transform of the spectrum that places
\(w_{p,b}\Psi(\mathbf h_b)\) on the beam grid. De-tilting removes the
parent carrier frequency \(\mathbf h_p\), enabling interpolation of the
slowly varying envelope across the aperture. This allows a coarse parent
set to represent the probe accurately.

In the full-parent limit (\(B_p=B\)), partitioned PRISM reduces to exact
PRISM (Section~\ref{sec-prism}; see Section~\ref{sec-supp-exactness} for
a term-by-term derivation). Partitioning thus introduces approximation
errors solely through the interpolation (Equation~\ref{eq-nnw}) of \(B\)
beams from \(B_p\) parents. For a Nyquist-step scan (integer
\(\symbf{\rho}\)), \(\widehat\psi_p(\mathbf r,\symbf{\rho})\) is a
cyclic shift of \(\widehat\psi_p(\mathbf r,\mathbf 0)\), which is
precomputed once, so each scan position needs only a roll and the
weighted sum over parents, skipping the per-position inverse FFT
entirely. Sub-pixel scans use the general path: a Fourier-shift ramp on
the beamlet spectrum and one inverse FFT per (scan, parent), chunked for
memory.

Memory. During the multislice advance, the memory-heavy phase, only the
\(B_p\) parent columns
\(\mathcal S\in\mathbb C^{B_p\times N_y\times N_x}\) are resident; the
de-tilt phase and beamlet basis are built lazily at probe-synthesis
time. The resident scattering-matrix footprint is thus reduced by a
factor of \(B/B_p\) compared to exact PRISM, providing a significant
memory advantage for large-scale simulations
(Section~\ref{sec-complexity}).

\paragraph{Magnitude-preserving reconstruction and focal
back-propagation}\label{sec-magnitude}

The de-tilting operation in Equation~\ref{eq-part-basis} removes each
parent's carrier frequency while retaining the parent-dependent phase
that the envelope \(\mathcal S^{\mathrm{dt}}_p\) acquires during
propagation. Consequently, the convex average (Equation~\ref{eq-nnw}) of
these complex envelopes decoheres:
\[\big|\textstyle\sum_p w_{p,b}\,\mathcal S^{\mathrm{dt}}_p\big| \le \sum_p
w_{p,b}\,\big|\mathcal S^{\mathrm{dt}}_p\big|.\] As a result, the
reconstructed columns, and thus the detector-integrated EELS signal,
systematically lose amplitude. This amplitude loss increases with
specimen thickness and is most pronounced for the adjoint detector
matrix \(\mathcal S_2\) (due to back-propagation through the remaining
thickness). At low parent counts, decoherence can suppress the absolute
map intensity by up to a factor of two, although the spatial pattern
remains intact.

To reduce this phase spread in elastic imaging, Pelz, Rakowski, et al.
(\citeproc{ref-pelz2022}{2021}) back-propagated the parent columns to
the probe-crossover (zero-defocus) plane before interpolation. This
approach applies to the probe-forming matrix \(\mathcal S_1\), but with
a key modification: since the convex natural-neighbor average is a
lossy, non-unitary mapping, the order of interpolation and propagation
affects the result. The de-tilted parents recohere most effectively
where their residual envelope phase is minimized---the scattering
centroid of the entrance-to-ionization path---rather than at the nominal
crossover. Interpolating at this centroid minimizes the error, which is
then preserved during the subsequent exact (unitary) propagation to the
ionization plane. We therefore back-propagate \(\mathcal S_1\) to this
centroid plane via a padded-window Fresnel step, interpolate, and then
propagate to the target plane. This procedure reduces the on-atom probe
reconstruction error by a factor of 2--3, sharpening the
\(\mathcal S_1\)-limited outputs (the 4D cube and qEELS; see
Section~\ref{sec-exp-regime}). This focal back-propagation is specific
to the probe leg: the detector matrix \(\mathcal S_2\) is a set of
parallel detector plane waves with no crossover plane (an empirical
back-propagation-distance sweep is flat), so the detector-summed map is
unaffected and remains governed by the \(\mathcal S_2\) reconstruction
(Section~\ref{sec-discussion}). Independently, we restore the lost
amplitude directly: the column magnitude is itself a convex
interpolation (a partition of unity over non-negative magnitudes, hence
decoherence-free), so we reconstruct

\begin{equation}\phantomsection\label{eq-magpreserve}{
\widetilde{\mathcal S}_b
  = \Big(\textstyle\sum_p w_{p,b}\,\big|\mathcal S^{\mathrm{dt}}_p\big|\Big)\,
    \frac{\sum_p w_{p,b}\,\mathcal S^{\mathrm{dt}}_p}
         {\big|\sum_p w_{p,b}\,\mathcal S^{\mathrm{dt}}_p\big|}\,
}\end{equation}

pairing the interpolated magnitude with the complex-sum phase (and
re-tilting as before). This is exact at the parents and reduces to exact
PRISM in the full-parent limit, so it adds no approximation there; at
finite \(B_p\) it recovers the correct absolute scale and restores
monotonic convergence of the map error in \(B_{p,1},B_{p,2}\)
(Section~\ref{sec-exp-accuracy}). Magnitude preservation is applied to
both \(\mathcal S_1\) and \(\mathcal S_2\) and costs one extra
window-sized reduction per atom; focal back-propagation is applied to
\(\mathcal S_1\) alone.

\begin{algorithm}[ht]
\caption{Partitioned PRISM-EELS (partition $\mathcal S_1$ only)}
\label{alg-partitioned}
\begin{algorithmic}[1]
\Require aperture $A$; transmissions $\{T_j\}$; propagator $\mathcal P$;
         ring counts $(n_{\text{radial}},n_{\text{angular}})$; $\{H_n\}$, sites, scans
\Statex Build (once per aperture; geometry only):
\State $\{\mathbf h_p\} \gets$ \textsc{HexRingParents}(aperture, $n_{\text{radial}},n_{\text{angular}}$) \Comment{$B_p$ beams}
\State $\mathbf w \gets$ \textsc{NaturalNeighborWeights}($\{\mathbf h_p\}, \{\mathbf h_b\}$) \Comment{cache, \eqref{eq-nnw}}
\State $\mathcal S_p \gets \mathcal F^{-1}[\delta_{\mathbf h_p}]$ for each parent
\Statex Per ionization plane $i$:
\State \textsc{AdvanceTo}($i$): apply \eqref{eq-smatrix-advance} to the $B_p$ parent columns
\State $\mathcal S^{\mathrm{dt}}_p \gets \mathcal S_p\cdot e^{-2\pi i\,\mathbf h_p\cdot\mathbf r/\mathbf N}$;\quad
       spectrum$[\mathbf h_b]\gets w_{p,b}\,\Psi(\mathbf h_b)$ \Comment{lazy, \eqref{eq-part-basis}}
\If{scan is integer (Nyquist step)}
  \State $\widehat\psi_p(\cdot,\mathbf 0) \gets \mathcal F^{-1}[\text{spectrum}]$ once;\quad
         $\psi(\cdot,\symbf{\rho})\gets\sum_p \mathrm{roll}(\widehat\psi_p,\symbf{\rho})\cdot\mathcal S^{\mathrm{dt}}_p$
\Else
  \State $\psi(\cdot,\symbf{\rho})\gets\sum_p \mathcal F^{-1}[\text{spectrum}\cdot e^{-2\pi i\,\mathbf q\cdot\symbf{\rho}/\mathbf N}]\cdot\mathcal S^{\mathrm{dt}}_p$ \Comment{sub-pixel}
\EndIf
\State accumulate exit intensities as in Alg.~\ref{alg-conventional}, lines 5--8
\end{algorithmic}
\end{algorithm}

\subsubsection{Dual-scattering-matrix PRISM-EELS for the elemental
map}\label{sec-dual}

The dominant STEM-EELS output is the elemental map \(I(\symbf{\rho})\),
Equation~\ref{eq-map}, the energy-filtered intensity collected over a
detector of semi-angle \(\theta_{\text{det}}\). Because the detector
accepts only the output beams \(\{\mathbf h_d\}_{d=1}^{n_{\text{det}}}\)
inside that aperture, the exit propagation collapses onto a second
scattering matrix \(\mathcal S_2\), the construction that makes the
FePt-nanoparticle-scale simulation of Brown, Ciston, and Ophus
(\citeproc{ref-brown2019}{2019}) tractable.

The detector exit matrix (reciprocity). The amplitude that an inelastic
wave \(w\) at the ionization plane delivers to detector beam \(d\) is
\(\langle\chi_{\mathbf h_d}\,|\,\mathcal M_{i\to\text{exit}}\,|\,w\rangle\).
By reciprocity this equals \(\langle \mathcal S_2^d\,|\,w\rangle\) where

\begin{equation}\phantomsection\label{eq-s2}{
\mathcal S_2^d \;=\; \mathcal M_{i\to\text{exit}}^{\dagger}\,\chi_{\mathbf h_d}
\qquad\Longleftrightarrow\qquad
\textstyle\sum_{\mathbf r}\mathcal S_2^d(\mathbf r)\,w(\mathbf r)
   = \big[\mathcal F\,\mathcal M_{i\to\text{exit}}\,w\big]_{\mathbf h_d},
}\end{equation}

i.e.~\(\mathcal S_2^d\) is the conjugate detector plane wave
\(\chi_{\mathbf h_d}^*=e^{-2\pi i\,\mathbf h_d\cdot\mathbf r/\mathbf N}\)
back-propagated from the exit surface to the ionization plane by
\emph{adjoint} (transpose) multislice, transmit at every slice and apply
the conjugate Fresnel step between slices, in reverse slice order. It is
built once from the exit surface and peeled forward one slice at a time,
so the same \(\mathcal S_2\) is reused across every atom in a slice
(and, being independent of \(\symbf{\rho}\), across every scan
position). As with \(\mathcal S_1\), the detector beams are
PRISM-subsampled by the interpolation factor so \(\mathcal S_2\) has
\(\sim\!f_yf_x\) fewer columns.

Localized transition coupling. Inserting the partitioned/PRISM probe
\(\psi=\sum_b c_b(\symbf{\rho})\,\mathcal S_1^b\), with illumination
coefficients
\(c_b(\symbf{\rho})=\Psi(\mathbf h_b)\,e^{-2\pi i\,\mathbf h_b\cdot\symbf{\rho}/\mathbf N}\),
the detector amplitude for channel \(n\), site \(\boldsymbol\tau\)
factorizes into a scan-independent transition coupling matrix and the
per-scan illumination:

\begin{equation}\phantomsection\label{eq-coupling}{
a_d(\symbf{\rho})=\sum_{b} M^{(n,\boldsymbol\tau)}_{d,b}\,c_b(\symbf{\rho}),
\qquad
M^{(n,\boldsymbol\tau)}_{d,b}
   =\!\!\sum_{\mathbf r\in\Omega_{\boldsymbol\tau}}\!\!
     \mathcal S_2^d(\mathbf r)\,H_{n0}(\mathbf r-\boldsymbol\tau)\,\mathcal S_1^b(\mathbf r).
}\end{equation}

Because \(H_{n0}\) is localized, the sum runs only over the small crop
window \(\Omega_{\boldsymbol\tau}\) (Section~\ref{sec-locality}). The
coupling matrix \(M^{(n,\boldsymbol\tau)}\) is therefore a small
\(n_{\text{det}}\times B\) matrix formed once per atom and channel, and
applied to all scan positions via a single matrix-vector product. The
map is

\begin{equation}\phantomsection\label{eq-map-sum}{
I(\symbf{\rho})=\sum_{\boldsymbol\tau}\sum_{n}\sum_{d}\big|a_d(\symbf{\rho})\big|^2 .
}\end{equation}

\(\mathcal S_1\) is built once and reused across scans; \(\mathcal S_2\)
is built once and reused across atoms; the per-atom work is a
crop-window contraction plus an inexpensive GEMM over scans. The
computational cost is therefore independent of the number of probe
positions \(P\) and scales linearly with the number of ionized atoms
(\citeproc{ref-brown2019}{Brown, Ciston, and Ophus 2019}). (A PRISM crop
mask restricts each atom's contribution to the scans inside its \(1/f\)
window.)

\begin{algorithm}[ht]
\caption{Double S-Matrix PRISM-EELS elemental map ($\mathcal S_1$ probe $+$ $\mathcal S_2$ detector)} [@brown2019]
\label{alg-dual}
\begin{algorithmic}[1]
\Require aperture $A$; transmissions $\{T_j\}$; $\{H_n\}$; sites; scans;
         detector semi-angle $\theta_{\text{det}}$; crop window $\Omega$
\State $\{\mathbf h_d\}\gets$ output beams with $|\mathbf q|\le\theta_{\text{det}}/\lambda$ (PRISM-subsampled)
\State build $\mathcal S_1$ (Alg.~2); build $\mathcal S_2$: $\mathcal S_2^d\gets$ adjoint-multislice of $\chi_{\mathbf h_d}^*$ from exit to slice $0$
\For{ionization plane $i$ with $\ge 1$ site}
  \State $\mathcal S_1.\textsc{AdvanceTo}(i)$;\quad $\mathcal S_2.\textsc{PeelTo}(i)$;\quad $c_b(\symbf{\rho})\gets$ illumination coeffs
  \For{each site $\boldsymbol\tau$ in slice $i$, each channel $n$}
    \State $M_{d,b}\gets\sum_{\mathbf r\in\Omega_{\boldsymbol\tau}}\mathcal S_2^d(\mathbf r)\,H_{n0}(\mathbf r-\boldsymbol\tau)\,\mathcal S_1^b(\mathbf r)$ \Comment{small, \eqref{eq-coupling}}
    \State $a_d(\symbf{\rho})\gets\sum_b M_{d,b}\,c_b(\symbf{\rho})$ for all scans $\symbf{\rho}$ \Comment{one GEMM}
    \State $I(\symbf{\rho})\mathrel{+}=\sum_d|a_d(\symbf{\rho})|^2$
  \EndFor
\EndFor
\end{algorithmic}
\end{algorithm}

\subsubsection{\texorpdfstring{Bi-partitioned PRISM-EELS (BiP-PRISM):
partition \(\mathcal S_1\) and
\(\mathcal S_2\)}{Bi-partitioned PRISM-EELS (BiP-PRISM): partition \textbackslash mathcal S\_1 and \textbackslash mathcal S\_2}}\label{sec-double}

The largest speed and memory improvements arise from partitioning both
scattering matrices. The construction of Section~\ref{sec-partitioned}
applies verbatim to \(\mathcal S_2\): by reciprocity the de-tilted
back-propagated detector columns are locally smooth across the
detector-beam set, so they are interpolated from a hex-ring subsample of
\(B_{p,2}\ll n_{\text{det}}\) parent detector beams with the same
natural-neighbor weights.

Windowed column reconstruction. Because the coupling
Equation~\ref{eq-coupling} is needed only on the crop window
\(\Omega_{\boldsymbol\tau}\), both matrices are reconstructed \emph{on
the window only}, never on the full grid: de-tilt the cropped parent
columns, NNW-combine to the full beam set, and re-tilt,

\begin{equation}\phantomsection\label{eq-window-recon}{
\mathcal S_1^b\big|_{\Omega}
  =\sum_{p} w^{(1)}_{b,p}\,\big(\mathcal S_1^{\mathrm{dt},p}\big|_{\Omega}\big)\,
     e^{2\pi i\,\mathbf h_b\cdot\mathbf r/\mathbf N},
\qquad
\mathcal S_2^d\big|_{\Omega}
  =\sum_{p} w^{(2)}_{d,p}\,\big(\mathcal S_2^{\mathrm{dt},p}\big|_{\Omega}\big)\,
     e^{2\pi i\,\mathbf h_d\cdot\mathbf r/\mathbf N}.
}\end{equation}

The persistent footprint is then just the \(B_{p,1}\) probe parents and
the \(B_{p,2}\) detector parents, so the resident scattering-matrix
memory drops from \(O\big((B+n_{\text{det}})\,G\big)\) to
\(O\big((B_{p,1}+B_{p,2})\,G\big)\), removing a practical bottleneck for
large grids, since \(\mathcal S_2\) storage is often dominant.

Exactness and error. Each partition is exact in its full-parent limit
(Equation~\ref{eq-part-exact}), so the double-partitioned map reduces to
the exact dual-matrix map (Section~\ref{sec-dual}) as
\(B_{p,1}\!\to\!B\), \(B_{p,2}\!\to\!n_{\text{det}}\). The two NNW
errors stack, but both are \emph{local}, the probe error on
\(\Omega_{\boldsymbol\tau}\) and the detector-column error on
\(\Omega_{\boldsymbol\tau}\), so the locality result
(Section~\ref{sec-locality}) bounds the map error by on-window
quantities and the operating point \(B_{p,1},B_{p,2}\) is chosen
accordingly.

\begin{algorithm}[ht]
\caption{BiP-PRISM EELS map (partition $\mathcal S_1$ and $\mathcal S_2$)}
\label{alg-double}
\begin{algorithmic}[1]
\Require as Alg.~4, plus parent counts for $\mathcal S_1$ and $\mathcal S_2$
\State build $\mathcal S_1$ on $B_{p,1}$ probe parents; build $\mathcal S_2$ on $B_{p,2}$ detector parents (adjoint multislice)
\State cache NNW weights $w^{(1)}$ (aperture) and $w^{(2)}$ (detector) \Comment{geometry only}
\For{ionization plane $i$ with $\ge 1$ site}
  \State $\mathcal S_1.\textsc{AdvanceTo}(i)$;\quad $\mathcal S_2.\textsc{PeelTo}(i)$
  \For{each site $\boldsymbol\tau$, each channel $n$}
    \State reconstruct $\mathcal S_1^b\big|_{\Omega_{\boldsymbol\tau}}$, $\mathcal S_2^d\big|_{\Omega_{\boldsymbol\tau}}$ from parents \Comment{\eqref{eq-window-recon}, window only}
    \State form $M_{d,b}$, apply to scans, accumulate $I$ \Comment{as Alg.~4, lines 6--8}
  \EndFor
\EndFor
\end{algorithmic}
\end{algorithm}

\subsubsection{Output modes: 4D cube and qEELS}\label{sec-outputs}

The three STEM-EELS outputs differ only in the final reduction over the
output-beam index \(d\) of the per-beam amplitudes \(a_d(\symbf{\rho})\)
from Equation~\ref{eq-coupling}:

\begin{itemize}
\item
  Elemental map, sum over all collected beams,
  \(I(\symbf{\rho})=\sum_{d}|a_d|^2\) (Equation~\ref{eq-map-sum});
  \(\mathcal S_2\) partitions (Section~\ref{sec-double}).
\item
  qEELS (momentum-resolved), keep one momentum axis and sum over the
  perpendicular one,
  \(I(q_\parallel,\symbf{\rho})=\sum_{d:\,h_{d,\perp}}|a_d|^2\); the
  output is one resolved diffraction axis per scan. \(\mathcal S_2\)
  still partitions, because its detector basis is NNW-reconstructed from
  parents regardless of how the beams are subsequently binned, provided
  the resolved axis is sampled at \(f=1\) and the detector-parent
  spacing along it sets the achievable \(q\)-resolution.
\item
  Energy-filtered 4D cube, keep every \(d\),
  \(I(\mathbf q,\symbf{\rho})=|a_{\mathbf q}|^2\). Full resolution
  requires the \emph{complete} output basis, so \(\mathcal S_2\) cannot
  be reduced; this is the \(\mathcal S_1\)-only regime of
  Section~\ref{sec-prism}--Section~\ref{sec-partitioned}, where the exit
  wave is propagated by ordinary multislice.
\end{itemize}

Thus qEELS sits between the fully-reduced map and the unreduced cube,
and the double-partitioned \(\mathcal S_1\)+\(\mathcal S_2\) machinery
covers the map and qEELS; only the cube forgoes the \(\mathcal S_2\)
win.

\subsection{Locality and memory complexity}\label{sec-locality}

\subsubsection{The locality result}\label{sec-locality-result}

Partitioned PRISM differs from exact PRISM in exactly one quantity: the
elastic probe \(\psi(\cdot,\symbf{\rho})\) delivered to the ionization
plane (Equation~\ref{eq-part-probe} vs. Equation~\ref{eq-prism-probe}).
Everything downstream in Equation~\ref{eq-cube}, the transition
potential \(H_{n0}\), the exit multislice
\(\mathcal M_{i\to\mathrm{exit}}\), and the final \(\mathcal F\), is
bit-for-bit identical between the two. We use this to bound the cube
error by a single, \emph{local} quantity.

Write the partitioned probe as \(\psi^{\mathrm{p}}=\psi+\delta\psi\),
where \(\psi\) is the exact probe and \(\delta\psi\) the interpolation
error from Equation~\ref{eq-nnw}. For a site \(\symbf{\tau}\) and
channel \(n\) the inelastic source Equation~\ref{eq-inelastic} differs
by

\begin{equation}\phantomsection\label{eq-loc-source}{
\delta\psi_n(\mathbf r)
  = H_{n0}(\mathbf r-\symbf{\tau})\,\delta\psi(\mathbf r).
}\end{equation}

Because \(H_{n0}\) is sharply localized about \(\symbf{\tau}\) (support
\(\Omega_{\symbf{\tau}}\)), \(\delta\psi_n\) depends on \(\delta\psi\)
only through its values on \(\Omega_{\symbf{\tau}}\):

\begin{equation}\phantomsection\label{eq-loc-bound}{
\|\delta\psi_n\|
  \;\le\; \big(\max_{\mathbf r}|H_{n0}(\mathbf r-\symbf{\tau})|\big)\,
          \big\|\delta\psi\big\|_{\Omega_{\symbf{\tau}}} .
}\end{equation}

The exit operator \(\mathcal M_{i\to\mathrm{exit}}\) is a product of
unitary transmission and Fresnel-propagation steps, and the detector
transform \(\mathcal F\) is unitary, so both are norm-preserving.
Therefore the error of the \emph{complex exit wavefield} equals the
error of its source,

\begin{equation}\phantomsection\label{eq-loc-unitary}{
\big\|\mathcal F\mathcal M_{i\to\mathrm{exit}}\,\delta\psi_n\big\|
  = \|\delta\psi_n\|,
}\end{equation}

and combining
Equation~\ref{eq-loc-bound}--Equation~\ref{eq-loc-unitary}, the
energy-filtered amplitude error contributed by each event is controlled
by the probe error at the ionized atom, and is independent of
\(\delta\psi\) elsewhere in the field of view. Summing incoherently over
sites and channels, the relative error of the diffraction intensity (the
4D cube) is found to be proportional to the on-atom probe error

\begin{equation}\phantomsection\label{eq-onatom}{
\varepsilon_{\text{on-atom}}
   \;\equiv\;
   \frac{\big\|\,\psi^{\mathrm{p}}-\psi\,\big\|_{\Omega_{\symbf{\tau}}}}
        {\big\|\,\psi\,\big\|_{\Omega_{\symbf{\tau}}}},
}\end{equation}

the quantity our experiments report alongside the cube error
(Section~\ref{sec-experiments}).

Partitioned PRISM maintains high accuracy when the natural-neighbor
interpolation reproduces the probe on the small region
\(\Omega_{\symbf{\tau}}\) near the ionized atom. This condition defines
the validity regime explored in Section~\ref{sec-experiments}. The
primary limitation is the specimen thickness: as electron channelling
redistributes intensity across the aperture, the on-atom probe becomes
more difficult to interpolate from a sparse set of parent beams, which
increases \(\varepsilon_{\text{on-atom}}\) and consequently the overall
simulation error. Deeper core-loss edges (corresponding to a narrower
\(H_{n0}\)) and moderate defocus values are well-tolerated
(Section~\ref{sec-exp-regime}). For in-focus probes and thin to
moderately thick specimens, the error remains low across a range of edge
energies. Strong defocus delocalizes the probe outside the region
\(\Omega_{\symbf{\tau}}\), making the locally-normalized
\(\varepsilon_{\text{on-atom}}\) metric less representative, although
the spatially integrated error remains robust to defocus. The same
support argument justifies evaluating the transition coupling
Equation~\ref{eq-coupling} only on a small window
\(\Omega_{\boldsymbol\tau}\) around each atom: \(H_{n0}\) vanishes
outside it, so the window-restricted coupling is exact up to the
(negligible) tail of \(H_{n0}\), independent of grid size. In the
double-partitioned map (Section~\ref{sec-double}) both scattering
matrices are reconstructed on this window only, and their two
interpolation errors, the probe error and the detector-column error, are
\emph{both} on-window quantities, so the locality result bounds the map
error by the same on-atom diagnostic and the operating point
\((B_{p,1},B_{p,2})\) is set by it.

\subsubsection{Complexity and memory}\label{sec-complexity}

Let \(G=N_yN_x\) be the grid size, \(P\) the number of scan positions,
\(N_Z\) the number of slices, and
\(N_{\mathrm{ev}}=\sum_{\symbf{\tau}}N_{\mathrm{ch}}\) the number of
(site \(\times\) channel) inelastic events. A single 2D FFT costs
\(O(G\log G)\). Table~\ref{tbl-complexity} collects the dominant terms.

\begin{longtable}[]{@{}
  >{\raggedright\arraybackslash}p{(\linewidth - 6\tabcolsep) * \real{0.1348}}
  >{\raggedright\arraybackslash}p{(\linewidth - 6\tabcolsep) * \real{0.2837}}
  >{\raggedright\arraybackslash}p{(\linewidth - 6\tabcolsep) * \real{0.2837}}
  >{\raggedright\arraybackslash}p{(\linewidth - 6\tabcolsep) * \real{0.2837}}@{}}
\caption{Dominant cost of the three algorithms. \(B\) = aperture beams,
\(B_p\) = parents (\(B_p\ll B\)). † the integer-scan roll path replaces
the per-position iFFT by a roll.}\label{tbl-complexity}\tabularnewline
\toprule\noalign{}
\begin{minipage}[b]{\linewidth}\raggedright
Stage
\end{minipage} & \begin{minipage}[b]{\linewidth}\raggedright
Conventional
\end{minipage} & \begin{minipage}[b]{\linewidth}\raggedright
PRISM (exact)
\end{minipage} & \begin{minipage}[b]{\linewidth}\raggedright
Partitioned
\end{minipage} \\
\midrule\noalign{}
\endfirsthead
\toprule\noalign{}
\begin{minipage}[b]{\linewidth}\raggedright
Stage
\end{minipage} & \begin{minipage}[b]{\linewidth}\raggedright
Conventional
\end{minipage} & \begin{minipage}[b]{\linewidth}\raggedright
PRISM (exact)
\end{minipage} & \begin{minipage}[b]{\linewidth}\raggedright
Partitioned
\end{minipage} \\
\midrule\noalign{}
\endhead
\bottomrule\noalign{}
\endlastfoot
Probe-forming (\(\mathcal S_1\)) build &
\begin{minipage}[t]{\linewidth}\raggedright
\begin{center}\rule{0.5\linewidth}{0.5pt}\end{center}
\end{minipage} & \(O(B\,N_Z\,G\log G)\) & \(O(B_p\,N_Z\,G\log G)\) \\
Probe synthesis (per plane) & \(O(P\,N_Z\,G\log G)\) & \(O(P\,B\,G)\) &
\(O(P\,B_p\,G)\) † \\
Exit propagation & \(O(P\,N_{\mathrm{ev}}\,N_Z\,G\log G)\) &
\(O(P\,N_{\mathrm{ev}}\,N_Z\,G\log G)\) &
\(O(P\,N_{\mathrm{ev}}\,N_Z\,G\log G)\) \\
Resident \(\mathcal S_1\) memory & \(O(G)\) & \(O(B\,G)\) &
\(O(B_p\,G)\) \\
\end{longtable}

These complexities highlight three important considerations:

\begin{enumerate}
\def\labelenumi{\arabic{enumi}.}
\item
  PRISM avoids the per-probe elastic calculation cost. The conventional
  elastic propagation step (row 2, \(O(P\,N_Z\,G\log G)\)) is replaced
  by a single \(\mathcal S_1\) calculation and a computationally
  inexpensive recombination step, resulting in the near-linear scaling
  described by (\citeproc{ref-brown2019}{Brown, Ciston, and Ophus
  2019}).
\item
  Beam partitioning reduces the \(\mathcal S_1\) calculation time and
  memory requirements by a factor of \(B/B_p\). Both the calculation
  step (row 1) and the resident matrix size (row 4) are decreased by
  this parent-reduction factor, which scales with both aperture size and
  field of view. This reduction provides the most significant
  performance improvements for computationally demanding simulations
  involving large grids, large apertures, and many potential slices.
\item
  The exit wave propagation remains a computational lower bound. The
  complexity of this step (row 3) is identical for all three methods and
  is unaffected by beam partitioning. The total reduction in simulation
  time is therefore limited by the proportion of time spent outside the
  exit propagation loop. While partitioning provides a \(B/B_p\) speedup
  for the \(\mathcal S_1\) calculation, the overall simulation speedup
  is constrained by the required exit propagation for each combination
  of scan position, channel, and atomic site.
  Section~\ref{sec-experiments} reports both the isolated
  \(\mathcal S_1\) calculation speedup and the total simulation speedup.
\end{enumerate}

\paragraph{The elemental map: removal of the exit
loop}\label{sec-complexity-map}

For the detector-integrated map, the exit propagation is no longer
computed per scan. Instead, \(\mathcal S_2\) is built once from the exit
surface and peeled forward, allowing reuse across all atoms. The
per-atom work consists of a crop-window contraction
(Equation~\ref{eq-coupling}) of complexity
\(O\big(n_{\text{det}}\,B\,|\Omega|\big)\) plus a General Matrix
Multiply (GEMM) of complexity
\(O\big(P\,n_{\text{det}}\,B_{\text{eff}}\big)\) over scan positions,
where \(|\Omega|=w_yw_x\ll G\) is the crop window. Consequently, the
\(O\big(P\,N_{\text{ev}}\,N_Z\,G\log G\big)\) exit multislice operation
in Table~\ref{tbl-complexity} is eliminated. The leading-order cost of
the map becomes independent of \(P\), achieving the linear scaling
described by Brown, Ciston, and Ophus (\citeproc{ref-brown2019}{2019}).
Table~\ref{tbl-map} summarizes the matrix calculation and memory
requirements.

\begin{longtable}[]{@{}
  >{\raggedright\arraybackslash}p{(\linewidth - 4\tabcolsep) * \real{0.2885}}
  >{\raggedright\arraybackslash}p{(\linewidth - 4\tabcolsep) * \real{0.3558}}
  >{\raggedright\arraybackslash}p{(\linewidth - 4\tabcolsep) * \real{0.3462}}@{}}
\caption{Elemental-map cost. Both matrix builds and the resident memory
shrink by the parent-reduction factors; the per-scan exit loop is absent
in both.}\label{tbl-map}\tabularnewline
\toprule\noalign{}
\begin{minipage}[b]{\linewidth}\raggedright
Quantity
\end{minipage} & \begin{minipage}[b]{\linewidth}\raggedright
Exact dual-matrix
\end{minipage} & \begin{minipage}[b]{\linewidth}\raggedright
Double-partitioned
\end{minipage} \\
\midrule\noalign{}
\endfirsthead
\toprule\noalign{}
\begin{minipage}[b]{\linewidth}\raggedright
Quantity
\end{minipage} & \begin{minipage}[b]{\linewidth}\raggedright
Exact dual-matrix
\end{minipage} & \begin{minipage}[b]{\linewidth}\raggedright
Double-partitioned
\end{minipage} \\
\midrule\noalign{}
\endhead
\bottomrule\noalign{}
\endlastfoot
\begin{minipage}[t]{\linewidth}\raggedright
Quantity
\end{minipage} & \begin{minipage}[t]{\linewidth}\raggedright
Exact dual-matrix
\end{minipage} & \begin{minipage}[t]{\linewidth}\raggedright
Double-partitioned
\end{minipage} \\
\(\mathcal S_1\) build & \(O(B\,N_Z\,G\log G)\) &
\(O(B_{p,1}\,N_Z\,G\log G)\) \\
\(\mathcal S_2\) build (adjoint) & \(O(n_{\text{det}}\,N_Z\,G\log G)\) &
\(O(B_{p,2}\,N_Z\,G\log G)\) \\
Resident \(\mathcal S_1+\mathcal S_2\) memory &
\(O\big((B+n_{\text{det}})\,G\big)\) &
\(O\big((B_{p,1}+B_{p,2})\,G\big)\) \\
Per-scan exit cost & --- (removed) & --- (removed) \\
\end{longtable}

Since storing \(\mathcal S_2\) (\(n_{\text{det}}\) detector columns)
typically dominates the memory footprint, partitioning this matrix
significantly relaxes the memory limits for large-grid simulations. For
the FePt-nanoparticle benchmark presented in
Section~\ref{sec-experiments}, measured on a GPU with 48 GB of RAM, the
exact dual-matrix simulation cannot be performed at the \(1908^2\) full
resolution, as the \(\mathcal S_2\) matrix alone requires approximately
19 GB of memory. In contrast, the double-partitioned simulation requires
only 12.6 GB of memory. For a \(936^2\) grid where both methods can be
computed, partitioning reduces the peak memory requirement from 12.7 to
2.6 GB (a factor of 4.8) and requires less calculation time, while
maintaining a Pearson cross-correlation of 0.996 relative to the exact
PRISM simulation.

\subsubsection{Measured scaling}\label{sec-scaling}

\begin{figure}[H]

\centering{

\includegraphics[width=1\linewidth,height=\textheight,keepaspectratio]{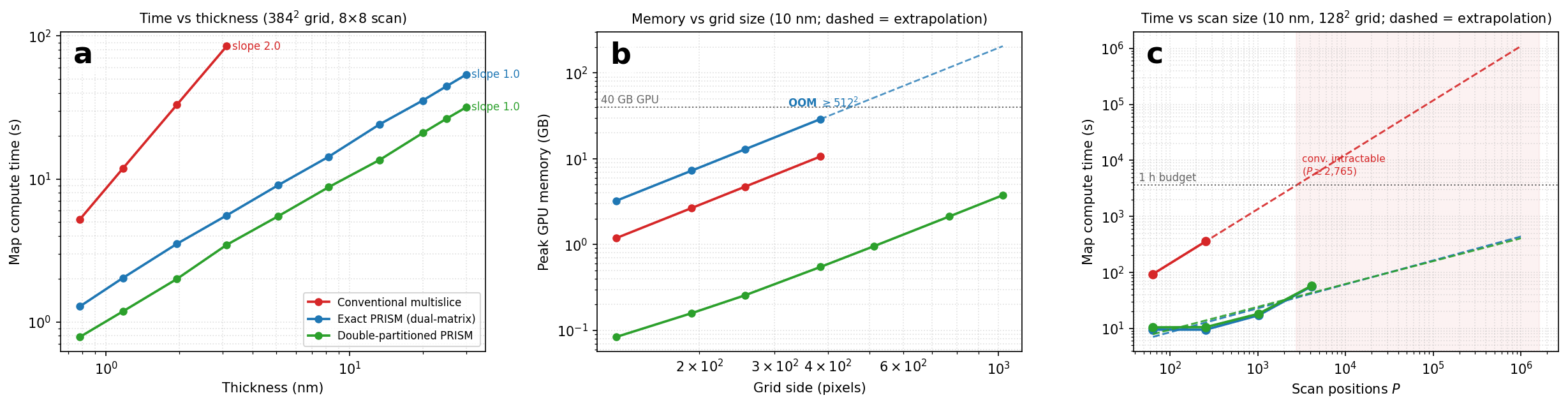}

}

\caption{\label{fig-scaling}Measured STEM-EELS map scaling comparisons
for conventional multislice, exact dual-matrix PRISM-EELS, and
double-partitioned PRISM-EELS on an NVIDIA A100 GPU (SrTiO\(_3\), Ti-L
edge, \((B_{p,1},B_{p,2})=(4,4)\)). (a) Compute time versus specimen
thickness: conventional multislice scales quadratically (\(O(N_Z^2)\)),
while both the exact and double-partitioned PRISM-EELS methods scale
linearly (\(O(N_Z)\)) by avoiding the per-scan exit wave propagation.
(b) Peak GPU memory versus grid side: the exact dual-matrix method
scales as \(O(n_{\det}G)\) and exceeds the 40 GB memory ceiling (dashed
line) for grids larger than \(384^2\), whereas the double-partitioned
method stays below 4 GB up to a \(1024^2\) grid. (c) Compute time versus
scan positions \(P\): conventional multislice scales as \(O(P)\) with a
large constant prefactor, becoming impractical beyond \(10^3\) positions
within an hour (shaded area), whereas the scattering matrix methods only
require a computationally efficient matrix multiplication, enabling over
\(10^5\) positions to be simulated.}

\end{figure}

Figure~\ref{fig-scaling} demonstrates the scaling trends outlined in
Table~\ref{tbl-scaling-setup} (Section~\ref{sec-supplementary}),
measured using an NVIDIA A100 (40 GB) GPU for a SrTiO\(_3\) slab at the
Ti-L edge with hydrogenic potentials. Three main scaling behaviors are
observed. (a) With specimen thickness, conventional multislice
computation time scales as \(O(N_Z^2)\) due to the exit multislice
operations required for each scan position and scattering event, which
both scale with thickness. In contrast, both scattering matrix methods
scale linearly (\(\sim\!N_Z\), consistent with Brown, Ciston, and Ophus
(\citeproc{ref-brown2019}{2019})), since the per-scan exit propagation
is eliminated. The double-partitioned simulation is approximately
\(1.7\times\) faster than the exact PRISM simulation, reflecting the
reduced computational cost of matrix construction and probe synthesis
(Section~\ref{sec-complexity-map}). (b) The peak memory requirements
show the most significant improvements from beam partitioning. The exact
dual-matrix \(\mathcal S_2\) memory usage scales as \(n_{\det}G\) and
exceeds the 40 GB memory limit for grids larger than \(384^2\)
(requiring 29 GB). Conversely, the double-partitioned method requires
less than 4 GB of memory up to a \(1024^2\) grid, is over \(50\times\)
more memory-efficient at \(384^2\), and is the only method capable of
scaling beyond a \(512^2\) grid on this hardware. (c) With the number of
scan positions \(P\), the conventional algorithm scales as \(O(P)\) with
a large constant factor (approximately 3 s per position in this example,
becoming impractical for more than \(10^3\) positions within an hour).
The scattering matrix methods only require the computationally efficient
\(O(P\,n_{\det}B_{\text{eff}})\) matrix multiplication step
(Section~\ref{sec-complexity-map}), which is two orders of magnitude
faster per position, allowing approximately \(10^5\) positions to be
simulated in the same time frame.

\section{Results}\label{sec-results}

\subsection{Numerical experiments}\label{sec-experiments}

We use conventional multislice (Section~\ref{sec-conventional}) as the
ground truth to benchmark the partitioned PRISM-EELS algorithm. The
exact (full-beam, full-detector) dual-matrix map and exact PRISM
(\(f=1\)) serve as the exact references. Transmission functions are
unitary (\(|T|=1\), representing no absorption) and each run uses a
single frozen-phonon configuration. Consequently, the timing is
independent of the phonon count, and the partitioning error is a
coherent, per-channel quantity. The radial backend is
\texttt{hydrogenic} wherever only the \emph{pattern} or the
\emph{timing} matters (both are backend-independent) and \texttt{gpaw}
where precise absolute edge energies/strengths are needed (as in the
LAO/STO showcase). GPU runs are on one RTX A6000 (48 GB); the qEELS demo
and the regime sweeps run on CPU at the small sizes shown. Per-figure
experimental setups are summarized in Table~\ref{tbl-setup}
(Section~\ref{sec-supplementary}); the map runtime/memory study of
Figure~\ref{fig-scaling} is set up in Table~\ref{tbl-scaling-setup}
(Section~\ref{sec-supplementary}).

\subsubsection{Accuracy vs.~parent counts}\label{sec-exp-accuracy}

Increasing the probe parents \(B_{p,1}\) and detector parents
\(B_{p,2}\) reduces the elemental-map error monotonically toward the
exact reference. This holds for both absolute (RMS) and pattern
(Pearson) metrics, provided the parent columns are recombined using the
magnitude-preserving natural-neighbor reconstruction
(Section~\ref{sec-magnitude}). While a naive complex average loses
coherent amplitude (underestimating the absolute intensity by up to a
factor of two at low parent counts), restoring the interpolated
magnitude recovers the correct intensity and converges to the exact
reference as the number of parents approaches the total beam count. The
useful operating point is the smallest \((B_{p,1},B_{p,2})\) that
satisfies the accuracy target, where the memory saving is largest.

\begin{figure}[H]

\centering{

\includegraphics[width=1\linewidth,height=\textheight,keepaspectratio]{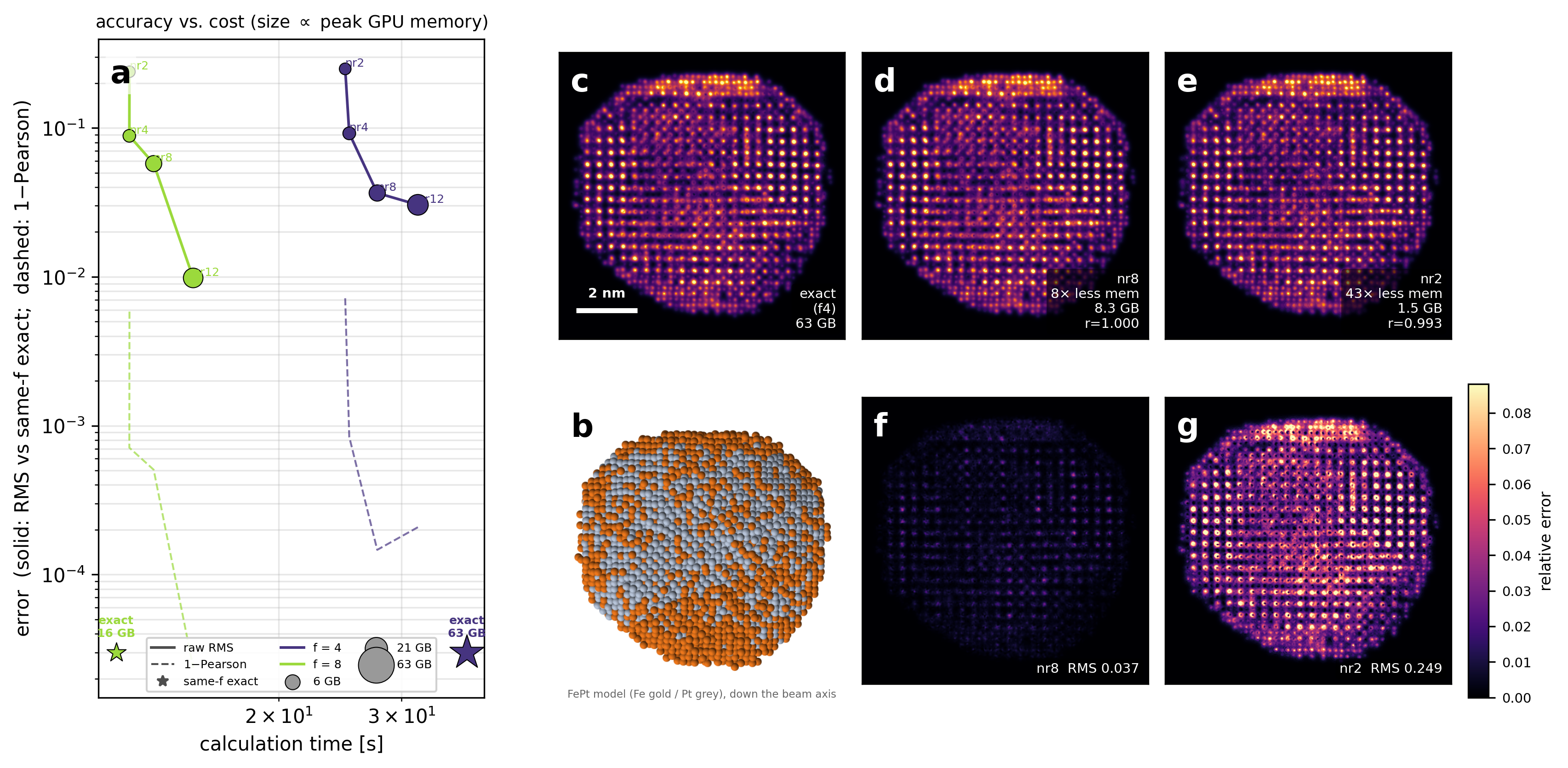}

}

\caption{\label{fig-accuracy-speed}Accuracy, speed, and memory of the
double-partitioned PRISM-EELS FePt Fe-L map vs.~parent count, at a
reduced \(936^2\) grid where the exact dual-matrix still fits (so it can
serve as the reference); at the \(1908^2\) resolution of
Table~\ref{tbl-fept} the exact dual-matrix exceeds GPU memory.}

\end{figure}

Figure~\ref{fig-accuracy-speed} sweeps the partition ring count (tying
\(B_{p,1}=B_{p,2}\) via \(n_{\text{radial}}\in\{2,4,8,12\}\)) at two
PRISM interpolation factors \(f\in\{4,8\}\) for the FePt Fe-L map
(\(936^2\) grid, \(8836\) scan positions, \(6569\) Fe), each compared
against the exact dual scattering matrix at the \emph{same} \(f\).
Partitioning shrinks the resident scattering matrices from \(63\) GB
(exact) to \(1.2\)--\(8\) GB (\(8\)--\(42\times\)) while preserving the
map: Pearson fidelity stays \(0.99\)--\(1.00\) across the sweep, and the
raw RMS error falls monotonically from \(\sim 25\%\) at
\(n_{\text{radial}}=2\) to a few percent by
\(n_{\text{radial}}=8\)--\(12\). At the \(n_{\text{radial}}=4\)
operating point the double-partitioned map matches the exact reference
to within \(\sim 9\%\) RMS error (Pearson correlation \(0.999\)) while
using \(\sim 22\times\) less memory.

\subsubsection{FePt nanoparticle: the speed and memory
benchmark}\label{sec-exp-fept}

We benchmark our method on the FePt nanoparticle Fe-L elemental map from
Brown, Ciston, and Ophus (\citeproc{ref-brown2019}{2019}), using the
same 23,196-atom structure (6,569 Fe sites), multislice parameters, and
Fe L-edge transition potentials. Double partitioning of both
\(\mathcal S_1\) and \(\mathcal S_2\) reduces peak GPU memory and matrix
construction time several-fold while retaining \(\sim\!0.99\) pattern
correlation with the exact dual-matrix map (Table~\ref{tbl-fept}; see
partitioned maps in Figure~\ref{fig-accuracy-speed}).

\begin{longtable}[]{@{}
  >{\raggedright\arraybackslash}p{(\linewidth - 8\tabcolsep) * \real{0.1875}}
  >{\raggedright\arraybackslash}p{(\linewidth - 8\tabcolsep) * \real{0.1875}}
  >{\raggedright\arraybackslash}p{(\linewidth - 8\tabcolsep) * \real{0.1875}}
  >{\raggedright\arraybackslash}p{(\linewidth - 8\tabcolsep) * \real{0.1875}}
  >{\raggedleft\arraybackslash}p{(\linewidth - 8\tabcolsep) * \real{0.2500}}@{}}
\caption{FePt Fe-L map (23196 atoms, 6569 Fe, 49 slices). At paper
resolution (\(1908^2\)) the exact dual-matrix exceeds the 48 GB GPU, its
\(\mathcal S_2\) build alone needs \(\sim\!19\) GB, while the
double-partitioned map runs at \(12.6\) GB. At a matched \(936^2\) where
both fit, partitioning cuts peak memory \(12.7\!\to\!2.6\) GB
(\(4.9\times\)) and is faster, at Pearson pattern correlation \(0.999\)
to the exact map; with the magnitude-preserving reconstruction
(Section~\ref{sec-magnitude}) the relative-\(L_2\) error is \(9.3\%\)
and the absolute scale is recovered to within \(\sim\!10\%\)
(\(\sim\!2\times\) too low without it).}\label{tbl-fept}\tabularnewline
\toprule\noalign{}
\begin{minipage}[b]{\linewidth}\raggedright
Configuration
\end{minipage} & \begin{minipage}[b]{\linewidth}\raggedright
grid
\end{minipage} & \begin{minipage}[b]{\linewidth}\raggedright
peak GPU mem
\end{minipage} & \begin{minipage}[b]{\linewidth}\raggedright
run time
\end{minipage} & \begin{minipage}[b]{\linewidth}\raggedleft
pattern fidelity
\end{minipage} \\
\midrule\noalign{}
\endfirsthead
\toprule\noalign{}
\begin{minipage}[b]{\linewidth}\raggedright
Configuration
\end{minipage} & \begin{minipage}[b]{\linewidth}\raggedright
grid
\end{minipage} & \begin{minipage}[b]{\linewidth}\raggedright
peak GPU mem
\end{minipage} & \begin{minipage}[b]{\linewidth}\raggedright
run time
\end{minipage} & \begin{minipage}[b]{\linewidth}\raggedleft
pattern fidelity
\end{minipage} \\
\midrule\noalign{}
\endhead
\bottomrule\noalign{}
\endlastfoot
exact dual-matrix & \(1908^2\) & OOM (\(>48\) GB) & --- & --- \\
double-partitioned & \(1908^2\) & \(12.6\,\text{GB}\) &
\(98.8\,\text{s}\) & --- \\
exact dual-matrix & \(936^2\) & \(12.7\,\text{GB}\) & \(26.4\,\text{s}\)
& \(1.000\) (ref.) \\
double-partitioned & \(936^2\) & \(2.6\,\text{GB}\) & \(22.8\,\text{s}\)
& \(0.999\) \\
\end{longtable}

\subsubsection{Validity regime and the locality
diagnostic}\label{sec-exp-regime}

\begin{figure}[H]

\centering{

\includegraphics[width=0.8\linewidth,height=\textheight,keepaspectratio]{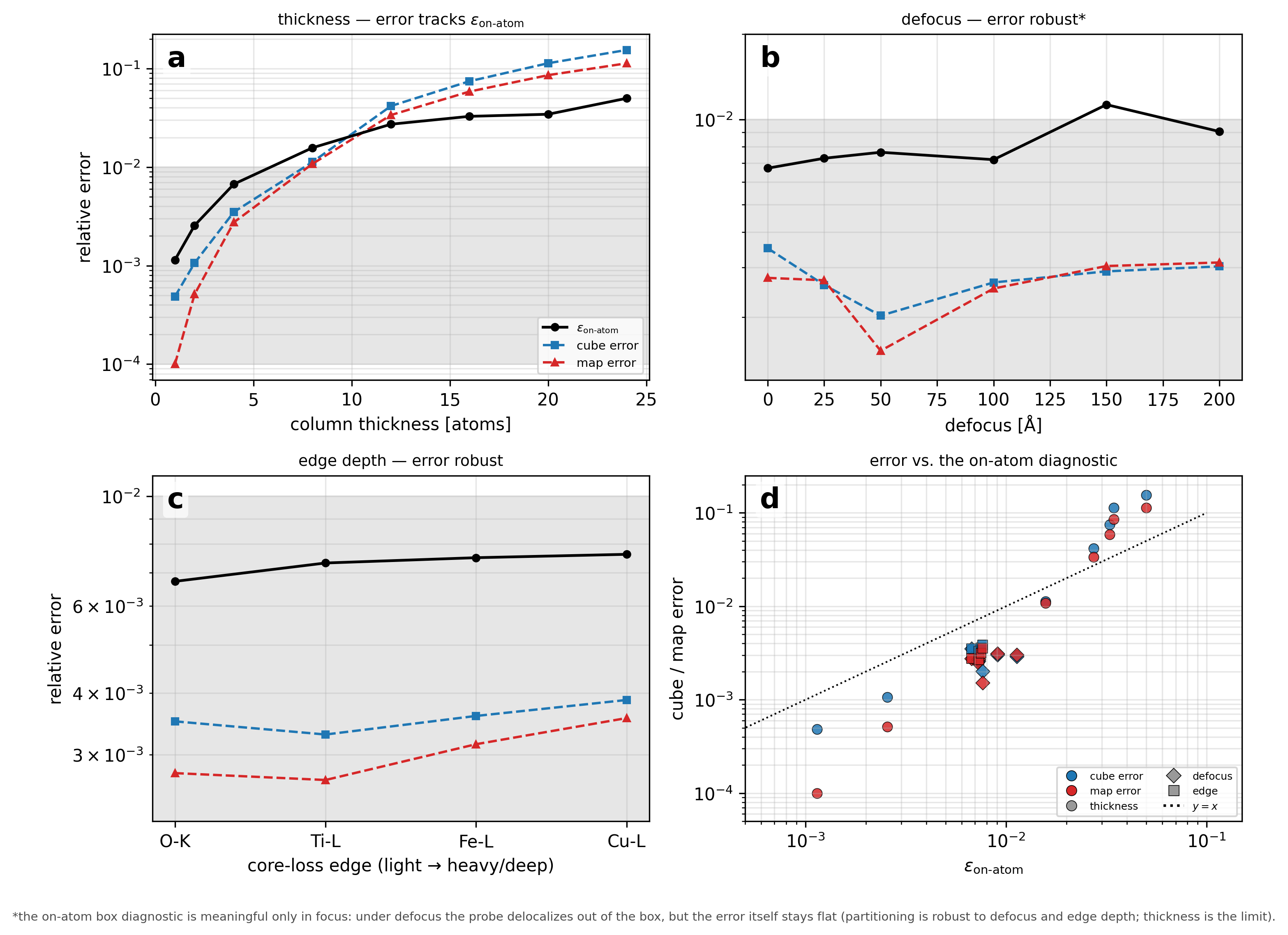}

}

\caption{\label{fig-regime}Validity regime of partitioned PRISM-EELS
(single column, \(n_{\text{radial}}=4\)). (a) Thickness; (b, c) defocus
and edge depth; (d) on-atom diagnostic vs.~error.}

\end{figure}

Figure~\ref{fig-regime} sweeps the three conditions that the locality
result (Section~\ref{sec-locality-result}) predicts could affect the
on-atom interpolation: thickness (channeling), defocus, and edge depth,
for a single atomic column (\(n_{\text{radial}}=4\), \(365\to61\) beams,
exact full-beam PRISM reference, GPAW transition potentials, with
\(\mathcal S_1\) focal back-propagation, Section~\ref{sec-magnitude}).
The dominant failure mode is specimen thickness. As channeling
redistributes intensity across the aperture, the partitioned cube and
map errors climb from under \(0.05\%\) to \(\sim 15\%\). The on-atom
diagnostic \(\varepsilon_{\text{on-atom}}\) rises in tandem (panel a),
demonstrating the locality result. In contrast, the method remains
robust to defocus and edge depth: the cube and map errors stay well
under \(1\%\) (often a few tenths of a percent) out to 200 Å defocus and
across the O-K to Cu-L series (panels b, c). Under strong defocus, the
probe delocalizes outside the fixed on-atom region, so
\(\varepsilon_{\text{on-atom}}\) (normalized by the vanishing in-box
probe norm) drifts upward as an artifact while the error itself stays
flat. STEM-EELS is typically operated in focus, where the diagnostic is
well-defined. Across the in-focus points, the error tracks
\(\varepsilon_{\text{on-atom}}\) (panel d), showing that the operating
point \((B_{p,1},B_{p,2})\) is governed by the on-atom term, with
thickness as the primary constraint.

\subsubsection{Multimodal showcase: LaAlO₃/SrTiO₃
interface}\label{sec-exp-laosto}

\begin{figure}[H]

\centering{

\includegraphics[width=1\linewidth,height=\textheight,keepaspectratio]{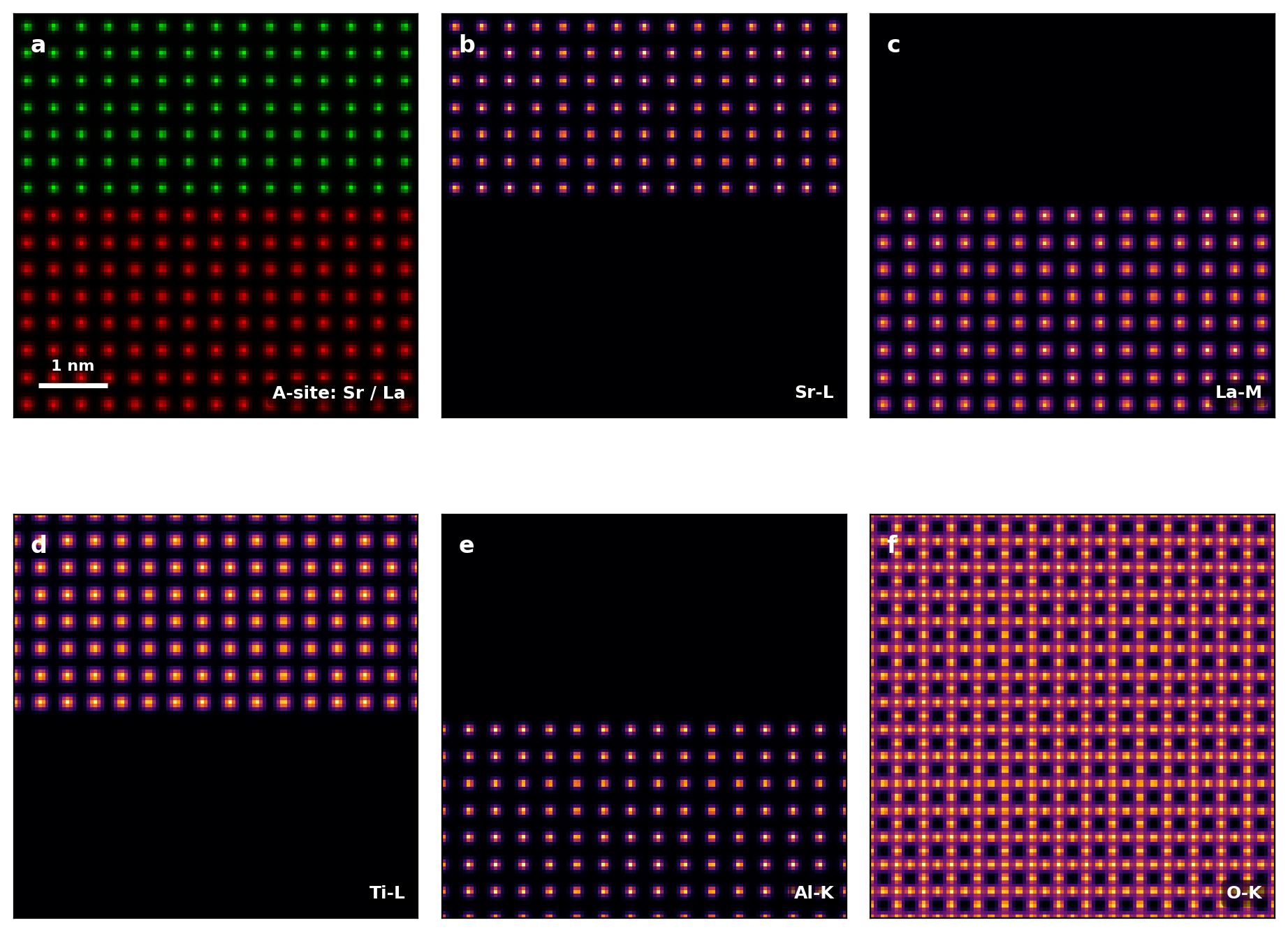}

}

\caption{\label{fig-laosto}LaAlO\(_3\)/SrTiO\(_3\) interface:
simultaneous atomic-resolution STEM-EELS maps (double-channeling,
double-partitioned PRISM; GPAW edges). (a) A-site Sr/La composite;
(b)--(f) Sr-L, La-M, Ti-L, Al-K, and O-K maps.}

\end{figure}

We demonstrate simultaneous atomic-resolution STEM-EELS maps of five
core-loss edges (Sr-L, La-M, Ti-L, Al-K, O-K) across a
LaAlO\(_3\)/SrTiO\(_3\) interface in cross section: each edge resolves
its sublattice (A-site Sr\(\to\)La, B-site Ti\(\to\)Al, anion O on both
sides), showing an atomically abrupt chemical step. The structure is a
\(16\!\times\!16\!\times\!6\)-cell film
(\(62\times62\times23\,\text{Å}\), \(640^2\) grid, 12 slices,
\(179^2=32041\) probe positions), with GPAW transition potentials and
double partitioning of both \(\mathcal S_1\) and \(\mathcal S_2\)
(\(n_{\text{radial}}=4\)). On one RTX A6000 the full five-edge run takes
\(\sim\!89\,\text{s}\) (Sr-L 14.6, La-M 37.2, Ti-L 14.4, Al-K 3.3, O-K
19.3 s, La-M carries 75 transition channels) at a peak GPU memory of
\(2.4\,\text{GB}\). The physical correctness of the underlying EELS
multislice model is validated against the reference package abTEM on a
line profile across the interface, yielding a Pearson correlation of
\(0.9986\) (see Section~\ref{sec-supp-validation} for details).
Figure~\ref{fig-laosto} resolves each sublattice and the abrupt A-site
Sr\(\to\)La step at the interface.

\subsubsection{Momentum-resolved qEELS}\label{sec-exp-qeels}

\begin{figure}[H]

\centering{

\includegraphics[width=1\linewidth,height=\textheight,keepaspectratio]{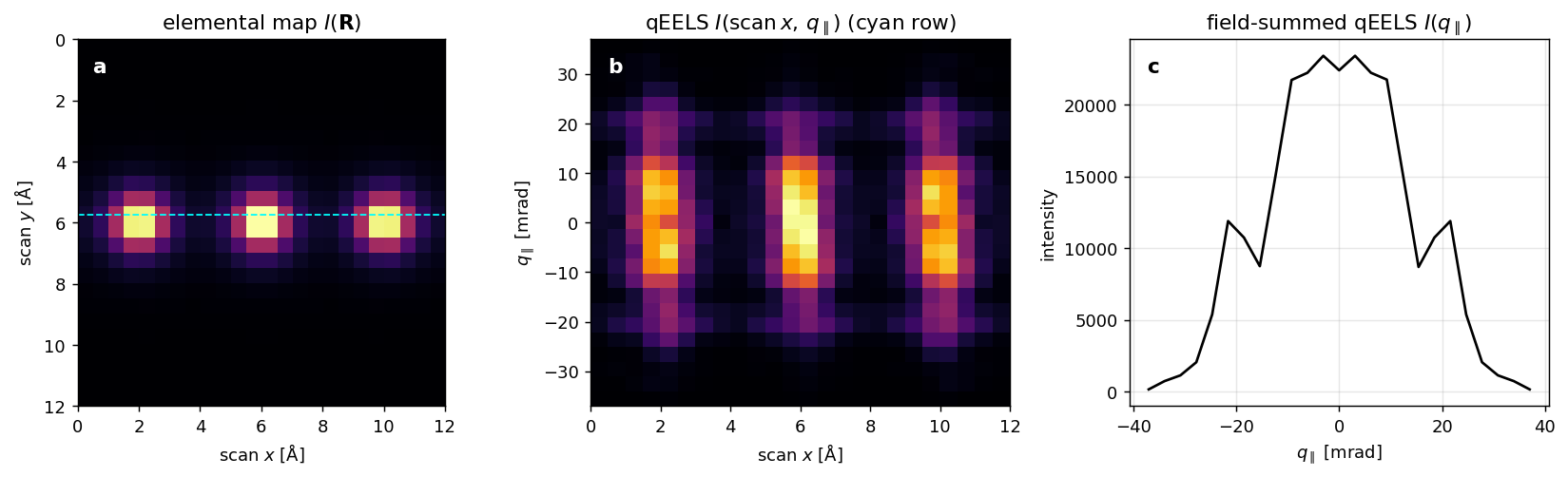}

}

\caption{\label{fig-qeels}Momentum-resolved (qEELS) output of
double-partitioned PRISM-EELS on a row of oxygen columns (O-K). (a)
Elemental map; (b) spectrum-image \(I(\mathrm{scan}\,x, q_\parallel)\)
along the highlighted row; (c) field-summed \(I(q_\parallel)\).}

\end{figure}

The double-partitioned matrices yield momentum-resolved (qEELS) output
by binning per-detector-beam intensities along one detector axis
(Section~\ref{sec-outputs}); summing the resolved axis recovers the
elemental map exactly, making this mode correct by construction without
adding scattering-matrix overhead. On a self-contained CPU demo
(hydrogenic O-K transition potentials, a row of oxygen columns, double
partitioning of \(\mathcal S_1\) and \(\mathcal S_2\)),
Figure~\ref{fig-qeels} shows the elemental map, the spectrum-image
\(I(\text{scan}\,x, q_\parallel)\) along the row, and the field-summed
\(I(q_\parallel)\); the sum-over-\(q_\parallel\) vs.~map residual is
\(\sim\!10^{-16}\) (the Section~\ref{sec-outputs} invariant).

\section{Discussion}\label{sec-discussion}

Because the partition weights for both matrices depend solely on the
aperture and detector geometries---independent of the specimen, defocus,
or thickness (Section~\ref{sec-partitioned},
Section~\ref{sec-double})---the parent counts \((B_{p,1},B_{p,2})\) can
be determined via an inexpensive pre-pass before committing to a full
simulation. The locality result (Section~\ref{sec-locality}) makes the
criterion concrete: the map and cube error are governed by the on-atom
reconstruction error \(\varepsilon_{\text{on-atom}}\), a single quantity
measurable from a handful of probe positions at one representative atom.
The user can increase the parent counts until
\(\varepsilon_{\text{on-atom}}\) meets the accuracy target; the smallest
counts \((B_{p,1},B_{p,2})\) that satisfy this target identify the most
memory-efficient operating point (Section~\ref{sec-exp-accuracy}). In
the full-parent limit the weights become the identity and exact PRISM is
recovered term-by-term (Equation~\ref{eq-part-exact}), so the
approximation can always be dialed back to the full PRISM algorithm.

The double-partitioned elemental map eliminates the per-scan inelastic
exit propagation entirely. The detector matrix \(\mathcal S_2\) is
constructed once and reused across all scan positions. This compresses
both resident matrices onto parent beams, reducing the peak GPU memory
footprint several-fold (from 12.7 to 2.6 GB, a \(4.9\times\) reduction,
for the FePt nanoparticle; Section~\ref{sec-exp-fept}). The memory
reduction is a decisive practical benefit: it renders tractable problems
that exceed GPU memory in the exact dual-matrix form (the
full-resolution FePt map, whose \(\mathcal S_2\) build alone needs
\(\sim\) 19 GB), and it scales with the aperture beam count \(B/B_p\),
so it grows with the large apertures and heterogeneous fields of view
that motivated linear-scaling PRISM-EELS
(\citeproc{ref-brown2019}{Brown, Ciston, and Ophus 2019}). In contrast,
the reduction in runtime per scan position is modest: the per-atom
window contraction is shared between the exact and partitioned paths, so
partitioning accelerates the matrix build specifically, and the
end-to-end map speedup is Amdahl-limited. The method is therefore
primarily an accelerator that enables memory feasibility.

A naive natural-neighbor reconstruction recombines the de-tilted parent
columns by a complex weighted average, which loses coherent amplitude
and underestimates the absolute map intensity by up to a factor of two
at low parent count. The magnitude-preserving reconstruction of
Section~\ref{sec-magnitude} removes this systematic deficit, restoring
the absolute scale to within \(\sim 10\%\) at the operating point
(relative-\(L_2\) error \(9.3\%\), Pearson \(0.999\) on FePt;
Section~\ref{sec-exp-fept}) while remaining exact in the full-parent
limit, so the partitioned map provides quantitative accuracy in addition
to pattern fidelity. The back-propagation-to-crossover remedy that
reduces the error in \emph{elastic} scattering-matrix imaging
(\citeproc{ref-pelz2022}{Pelz, Rakowski, et al. 2021}) does carry over,
but only to the probe leg \(\mathcal S_1\): since the convex NNW average
is lossy, the order of propagation and interpolation affects the result.
Consequently, back-propagating \(\mathcal S_1\) to the
scattering-centroid plane before interpolation minimizes the error,
which persists after exact re-propagation to the ionized-atom plane.
This procedure reduces the on-atom probe error by a factor of 2--3,
improving the \(\mathcal S_1\)-limited cube and qEELS outputs
(Section~\ref{sec-magnitude}). This technique is inapplicable to the
detector matrix \(\mathcal S_2\), whose parallel plane waves have no
crossover, so the detector-summed map remains \(\mathcal S_2\)-limited.
This bottleneck exhibits a structure we have not yet exploited. Summing
the detector index collapses the map contribution of \(\mathcal S_2\) to
a per-slice response operator whose effective rank on the inelastic
window is small (\(\sim\!40\) modes versus several hundred beams).
Consequently, the map's detector contraction is in principle a
low-dimensional, per-slice \emph{transfer function}. This offers a
potential route to a faster map (and a controllable speed/accuracy
trade-off) that we leave to future work, noting that the coherent
dynamical response cannot be reduced to a single propagated intensity.

The elemental map, momentum-resolved qEELS, and the full 4D cube are a
single computation differing only in the reduction taken over the
detector-beam index (Section~\ref{sec-outputs}): the map sums all
detector beams, qEELS resolves one momentum axis, and the cube retains
every beam. Double partitioning accelerates the first two; the cube is
the one mode in which \(\mathcal S_2\) cannot be reduced, because no
detector sum is taken, so there only \(\mathcal S_1\) is partitioned.
For qEELS the detector-parent spacing sets a momentum-resolution floor,
a trade-off between \(q\)-resolution and the \(B_{p,2}\) memory budget,
that the same on-atom criterion governs.

The locality result allows a single measurable check, and the regime
survey (Section~\ref{sec-exp-regime}) shows the answer is encouraging:
the binding constraint is specimen \emph{thickness}, through which
channelling redistributes intensity across the aperture and raises
\(\varepsilon_{\text{on-atom}}\), while the method is robust to defocus
and to edge depth across the O-K\(\to\)Cu-L range. Since
atomic-resolution STEM-EELS is acquired in focus, the practically
relevant axis is thickness, and there the on-atom diagnostic both
predicts the onset of error and prescribes the parent counts needed to
suppress it. For very thick specimens one simply raises
\((B_{p,1},B_{p,2})\) toward the exact limit.

The method combines and extends three threads: beam-subsampling
interpolation of the scattering matrix (\citeproc{ref-ophus2017}{Ophus
2017}), its \emph{partitioning} onto parent beams for elastic imaging
(Pelz, Brown, et al. (\citeproc{ref-pelz2021}{2021}); Pelz, Rakowski, et
al. (\citeproc{ref-pelz2022}{2021})), and the linear-scaling reuse of a
probe-forming matrix across scan positions for inelastic imaging
(\citeproc{ref-brown2019}{Brown, Ciston, and Ophus 2019}). Our
contribution is to partition \emph{both} the probe-forming and the
adjoint detector matrix and to make the inelastic reconstruction
quantitative. This is complementary to parallelization strategies that
target the stochastic inelastic average, such as the phonon/plasmon
phase-scrambling of Mendis (\citeproc{ref-mendis2023}{2023}), and to the
transition-potential multislice codes it benchmarks against
(\citeproc{ref-allen2015}{Allen, D'Alfonso, and Findlay 2015}). Finally,
the scattering matrix the method computes is the same object
reconstructed from 4D-STEM data by ptychography
(\citeproc{ref-findlay2021}{Findlay et al. 2021};
\citeproc{ref-pelz2021}{Pelz, Brown, et al. 2021};
\citeproc{ref-terzoudis2023}{Terzoudis-Lumsden et al. 2023}); an
efficient forward model and an experimental estimate of the same matrix
are natural candidates for simulation-constrained reconstruction.

The core-loss filtered 4D-STEM mode retains the full per-scan exit
computation on the \(\mathcal S_1\) side, so its gains are smaller than
the map's; reducing this exit-side cost, for instance a small-detector
\(\mathcal S_2\) for non-energy-filtered signals, or a restricted exit
propagation, is the clearest avenue for further acceleration. The
present results use a single frozen-phonon configuration; combining
partitioning with frozen-phonon/quantum-excitation averaging and
multi-edge batching, both of which reuse the same elastic matrices, is
straightforward and would amortize the build further. The memory
reduction makes GPU-resident simulation of large, thick, heterogeneous
cells practical, opening the door to first-principles transition
potentials (\citeproc{ref-madsen2021}{Madsen and Susi 2021}) and to the
momentum-resolved and vibrational EELS regimes
(\citeproc{ref-krivanek2014}{Krivanek et al. 2014};
\citeproc{ref-hage2018}{Hage et al. 2018}) where the qEELS output mode
is most valuable.

\section{Conclusion}\label{sec-conclusion}

We have presented an algorithm that partitions both PRISM-EELS
scattering matrices: the probe-forming matrix \(\mathcal S_1\) and the
adjoint detector matrix \(\mathcal S_2\). Each matrix is calculated on a
reduced set of parent beams, and the full aperture is reconstructed
using magnitude-preserving natural-neighbor interpolation within a
localized window around the ionized atom. For detector-integrated
elemental mapping, this method eliminates the per-scan exit wave
propagation and significantly reduces resident GPU memory requirements.
For a 23,000-atom FePt nanoparticle simulation, the memory drops from
12.7 to 2.6 GB, achieving a Pearson correlation of 0.999 and a 9\%
relative \(L_2\) error compared to the exact PRISM simulation. These
memory improvements allow full-resolution maps to be calculated in cases
where the exact dual-matrix simulation exceeds the available GPU memory.
The momentum-resolved qEELS and 4D-cube signals can also be calculated
as reductions over the detector beams. We demonstrated the utility of
this method for large fields of view by simulating a five-edge
oxide-interface map, which required a peak memory of only 2.4 GB. We
have also shown that the approximation error is localized to the on-atom
reconstruction, which effectively predicts the validity regime of the
algorithm. This regime is primarily limited by specimen thickness, while
remaining robust to variations in defocus and edge energy. Consequently,
the optimal simulation parameters can be determined by evaluating the
error of the on-atom probe. Since the partitioned method converges to
exact PRISM as the number of parent beams approaches the total beam
count, users can balance computational efficiency and memory usage
against simulation accuracy using a controllable error metric.

\section*{Code and data availability}\label{code-and-data-availability}
\addcontentsline{toc}{section}{Code and data availability}

Code to reproduce the results will be made available upon publication at
https://github.com/scatterem.

\section*{Acknowledgments}\label{acknowledgments}
\addcontentsline{toc}{section}{Acknowledgments}

This work received funding from the European Research Council (ERC)
under the Horizon Europe research and innovation programme (grant
agreement No.~101164581, project HyperScaleEM) and from the Deutsche
Forschungsgemeinschaft (DFG, German Research Foundation) through the
Research Training Group GRK 3103 \emph{CorMic: Korrelative
Materialmikroskopie -- Von nanostrukturierten funktionalen Filmen zu
hierarchischen Funktionsmaterialien} (project number 537140136).

The author thanks the developers of the muSTEM,PRISM, Prismatic,
py\_multislice and abTEM projects, on whose ideas this work builds.

\section*{Conflict of interest}\label{conflict-of-interest}
\addcontentsline{toc}{section}{Conflict of interest}

The author declares no conflicts of interest.

\section*{Ethics statement}\label{ethics-statement}
\addcontentsline{toc}{section}{Ethics statement}

This work is entirely computational and did not involve any human
participants or animal subjects. No ethical approval was required.

\section{Notation}\label{sec-notation}

\begin{longtable}[]{@{}
  >{\raggedright\arraybackslash}p{(\linewidth - 2\tabcolsep) * \real{0.3717}}
  >{\raggedright\arraybackslash}p{(\linewidth - 2\tabcolsep) * \real{0.6283}}@{}}
\caption{Symbols used in the text.}\label{tbl-notation}\tabularnewline
\toprule\noalign{}
\endfirsthead
\endhead
\bottomrule\noalign{}
\endlastfoot
\begin{minipage}[t]{\linewidth}\raggedright
Symbol
\end{minipage} & \begin{minipage}[t]{\linewidth}\raggedright
Meaning
\end{minipage} \\
\(N_y, N_x\); \(\mathbf N\) & grid pixels; the pair \((N_y,N_x)\) \\
\(G=N_yN_x\) & total grid size \\
\(N_Z\), \(\Delta z\) & number of slices; inter-slice distance \\
\(\mathbf r,\ \mathbf q,\ \mathbf h\) & real-space coord., spatial
frequency, integer beam index \\
\(\symbf{\rho}\), \(P\) & scan position (pixels); number of scan
positions \\
\(\Psi(\mathbf h)\), \(c_b\) & probe-forming aperture; coefficient
\(\Psi(\mathbf h_b)\) of beam \(b\) \\
\(B\), \(\{\mathbf h_b\}\) & number of aperture beams; their indices \\
\(B_p\), \(\{\mathbf h_p\}\) & number of parent beams; their indices \\
\(f_y,f_x\) & PRISM interpolation (beam-subsampling) factors \\
\(T_j(\mathbf r)\), \(\sigma\), \(V_j\) & slice transmission;
interaction constant; projected potential \\
\(\mathcal P\), \(P(\mathbf q)\), \(\lambda\) & Fresnel propagator
operator / transfer function; wavelength \\
\(\mathcal F\) & 2D discrete Fourier transform (detector transform) \\
\(\psi_0,\ \psi\) & incident probe; elastic wave at the ionization
plane \\
\(H_{n0}\), \(N_{\mathrm{ch}}\) & transition potential for channel
\(n\); number of channels \\
\(\symbf{\tau}\), \(\Omega_{\symbf{\tau}}\) & ionized-atom position;
support of \(H_{n0}\) about it \\
\(\mathcal M_{i\to\mathrm{exit}}\) & multislice exit operator from slice
\(i\) \\
\(I(\mathbf q,\symbf{\rho})\) & energy-filtered diffraction intensity
(the 4D cube) \\
\(\mathcal S_1\), \(\mathcal S_1^b\) & probe-forming scattering matrix;
its row for beam \(b\) \\
\(\mathcal S_2\), \(\mathcal S_2^d\) & detector exit matrix;
adjoint-multislice column for detector beam \(d\) \\
\(\mathcal S^{\mathrm{dt}}_p\) & de-tilted parent column \(p\) \\
\(\theta_{\text{det}}\), \(\mathbf h_d\), \(n_{\text{det}}\) & detector
collection semi-angle; detector beams; their count \\
\(B_{p,1}\), \(B_{p,2}\) & probe parents (\(\equiv B_p\)) and detector
parents \\
\(c_b(\symbf{\rho})\) & illumination coefficient
\(\Psi(\mathbf h_b)e^{-2\pi i\mathbf h_b\cdot\symbf{\rho}/\mathbf N}\) \\
\(M^{(n,\boldsymbol\tau)}_{d,b}\), \(a_d(\symbf{\rho})\) & transition
coupling matrix; detector-beam amplitude \\
\(I(\symbf{\rho})\), \(I(q_\parallel,\symbf{\rho})\) & elemental map;
momentum-resolved (qEELS) output \\
\(\mathbf w\), \(w_{p,b}\) & natural-neighbor weight matrix
(\(B_p\times B\)) and entries \\
\(\widehat\psi_p\) & beamlet basis for parent \(p\)
(\citeproc{ref-pelz2022}{Pelz, Rakowski, et al. 2021}, Eq. 17)
\textbar{} \\
\(n_{\text{radial}}, n_{\text{angular}}\) & hex-ring parent-sampling
counts \\
\(\varepsilon_{\text{on-atom}}\) & relative probe error on
\(\Omega_{\symbf{\tau}}\) (locality diagnostic) \\
\end{longtable}

\section{Supplementary Material}\label{sec-supplementary}

\subsection{Per-figure experimental setups}\label{sec-supp-setups}

\begin{longtable}[]{@{}
  >{\raggedright\arraybackslash}p{(\linewidth - 18\tabcolsep) * \real{0.0909}}
  >{\raggedright\arraybackslash}p{(\linewidth - 18\tabcolsep) * \real{0.0909}}
  >{\raggedleft\arraybackslash}p{(\linewidth - 18\tabcolsep) * \real{0.1212}}
  >{\raggedright\arraybackslash}p{(\linewidth - 18\tabcolsep) * \real{0.0909}}
  >{\raggedright\arraybackslash}p{(\linewidth - 18\tabcolsep) * \real{0.0909}}
  >{\raggedright\arraybackslash}p{(\linewidth - 18\tabcolsep) * \real{0.0909}}
  >{\raggedright\arraybackslash}p{(\linewidth - 18\tabcolsep) * \real{0.0909}}
  >{\raggedright\arraybackslash}p{(\linewidth - 18\tabcolsep) * \real{0.0909}}
  >{\raggedleft\arraybackslash}p{(\linewidth - 18\tabcolsep) * \real{0.1212}}
  >{\raggedleft\arraybackslash}p{(\linewidth - 18\tabcolsep) * \real{0.1212}}@{}}
\caption{Per-figure experimental setup (rows top-to-bottom correspond to
the accuracy/parents, FePt, regime, LAO/STO, qEELS, and abTEM validation
figures in Section~\ref{sec-experiments} and
Section~\ref{sec-supplementary}). \(\alpha\) = probe convergence
semi-angle, \(\theta_{\det}\) = EELS collection semi-angle (``full'' =
un-integrated 4D cube, no detector matrix). \((B_{p,1},B_{p,2})\) =
probe / detector parent counts (\(n_{\text{radial}}=4\) unless swept;
``\(-\)'' = that matrix not partitioned). PRISM interpolation factor and
crop window: \(f=9\), \(|\Omega|\approx86\) px (FePt); \(f=4\), 22 px
(LAO/STO); \(f=1\), full window elsewhere. Continuum energy above
threshold \(\varepsilon=25\) eV (FePt), 5 eV (LAO/STO). The
accuracy/regime experiments report the 4D-cube error (full detector);
all map figures integrate to the collection
semi-angle.}\label{tbl-setup}\tabularnewline
\toprule\noalign{}
\begin{minipage}[b]{\linewidth}\raggedright
Experiment
\end{minipage} & \begin{minipage}[b]{\linewidth}\raggedright
System
\end{minipage} & \begin{minipage}[b]{\linewidth}\raggedleft
\(E_0\)/kV
\end{minipage} & \begin{minipage}[b]{\linewidth}\raggedright
grid \(G\)
\end{minipage} & \begin{minipage}[b]{\linewidth}\raggedright
\(\alpha\,/\,\theta_{\det}\) /mrad
\end{minipage} & \begin{minipage}[b]{\linewidth}\raggedright
slices (\(\Delta z\))
\end{minipage} & \begin{minipage}[b]{\linewidth}\raggedright
edge(s)
\end{minipage} & \begin{minipage}[b]{\linewidth}\raggedright
backend
\end{minipage} & \begin{minipage}[b]{\linewidth}\raggedleft
\((B_{p,1},B_{p,2})\)
\end{minipage} & \begin{minipage}[b]{\linewidth}\raggedleft
scan \(P\)
\end{minipage} \\
\midrule\noalign{}
\endfirsthead
\toprule\noalign{}
\begin{minipage}[b]{\linewidth}\raggedright
Experiment
\end{minipage} & \begin{minipage}[b]{\linewidth}\raggedright
System
\end{minipage} & \begin{minipage}[b]{\linewidth}\raggedleft
\(E_0\)/kV
\end{minipage} & \begin{minipage}[b]{\linewidth}\raggedright
grid \(G\)
\end{minipage} & \begin{minipage}[b]{\linewidth}\raggedright
\(\alpha\,/\,\theta_{\det}\) /mrad
\end{minipage} & \begin{minipage}[b]{\linewidth}\raggedright
slices (\(\Delta z\))
\end{minipage} & \begin{minipage}[b]{\linewidth}\raggedright
edge(s)
\end{minipage} & \begin{minipage}[b]{\linewidth}\raggedright
backend
\end{minipage} & \begin{minipage}[b]{\linewidth}\raggedleft
\((B_{p,1},B_{p,2})\)
\end{minipage} & \begin{minipage}[b]{\linewidth}\raggedleft
scan \(P\)
\end{minipage} \\
\midrule\noalign{}
\endhead
\bottomrule\noalign{}
\endlastfoot
Accuracy--parents & O column (1 site) & 100 & \(256^2\) &
\(40\,/\,\)full & 32 & O-K & hydrog. & \(B_{p,1}\!\in\!\{4..32\}\) &
\(4^2\) \\
FePt map & FePt NP, 23196 atoms (6569 Fe) & 300 & \(1908^2\) (\(936^2\))
& \(20\,/\,20\) & 49 (2 Å) & Fe-L & hydrog. & \((4,4)\) &
\(\sim\!35000\) \\
Validity regime & O / Ti / Cu columns & 100 & \(128^2\!/\,256^2\) &
\(35\,/\,25\), full & 5--120 (0.5 Å) & O-K, Ti-L, Cu-L & gpaw / hydrog.
& \((4,-)\) & \(3^2\) \\
LAO/STO & LaAlO₃/SrTiO₃, \(16\!\times\!16\!\times\!6\) cells & 300 &
\(640^2\) & \(20\,/\,20\) & 12 (2/cell) & Sr-L, La-M, Ti-L, Al-K, O-K &
gpaw & \((4,4)\) & \(179^2\) \\
qEELS & O-column row & 100 & \(96^2\) & \(20\,/\,40\) & 4 & O-K &
hydrog. & \((4,4)\) & \(24^2\) \\
Validation & LaAlO₃/SrTiO₃ line scan & 300 & \(384\times32\) &
\(20\,/\,20\) & 4 (1.95 Å) & Ti-L & gpaw & exact (1,1) & 94 \\
\end{longtable}

\subsection{Map-scaling study setup}\label{sec-supp-scaling}

\begin{longtable}[]{@{}
  >{\raggedright\arraybackslash}p{(\linewidth - 10\tabcolsep) * \real{0.1500}}
  >{\raggedright\arraybackslash}p{(\linewidth - 10\tabcolsep) * \real{0.1500}}
  >{\raggedright\arraybackslash}p{(\linewidth - 10\tabcolsep) * \real{0.1500}}
  >{\raggedright\arraybackslash}p{(\linewidth - 10\tabcolsep) * \real{0.1500}}
  >{\raggedleft\arraybackslash}p{(\linewidth - 10\tabcolsep) * \real{0.2000}}
  >{\raggedleft\arraybackslash}p{(\linewidth - 10\tabcolsep) * \real{0.2000}}@{}}
\caption{Setup for the map-scaling study of Figure~\ref{fig-scaling}
(Section~\ref{sec-scaling}). All panels: \(100\) kV, \(\alpha=30\) /
\(\theta_{\det}=50\) mrad, Ti-L edge, hydrogenic backend,
double-partitioned \((B_{p,1},B_{p,2})=(4,4)\), crop window \(f=16\),
\texttt{contract="probe"}, one NVIDIA A100 (40 GB). Conventional
multislice is plotted only where tractable (capped at \(\approx\!3\) nm
in (a) and \(P\le256\) in (c); beyond that its \(O(N_Z^2)\) / \(O(P)\)
cost is prohibitive).}\label{tbl-scaling-setup}\tabularnewline
\toprule\noalign{}
\begin{minipage}[b]{\linewidth}\raggedright
Panel
\end{minipage} & \begin{minipage}[b]{\linewidth}\raggedright
swept
\end{minipage} & \begin{minipage}[b]{\linewidth}\raggedright
system
\end{minipage} & \begin{minipage}[b]{\linewidth}\raggedright
grid \(G\)
\end{minipage} & \begin{minipage}[b]{\linewidth}\raggedleft
slices \(N_Z\)
\end{minipage} & \begin{minipage}[b]{\linewidth}\raggedleft
scan \(P\)
\end{minipage} \\
\midrule\noalign{}
\endfirsthead
\toprule\noalign{}
\begin{minipage}[b]{\linewidth}\raggedright
Panel
\end{minipage} & \begin{minipage}[b]{\linewidth}\raggedright
swept
\end{minipage} & \begin{minipage}[b]{\linewidth}\raggedright
system
\end{minipage} & \begin{minipage}[b]{\linewidth}\raggedright
grid \(G\)
\end{minipage} & \begin{minipage}[b]{\linewidth}\raggedleft
slices \(N_Z\)
\end{minipage} & \begin{minipage}[b]{\linewidth}\raggedleft
scan \(P\)
\end{minipage} \\
\midrule\noalign{}
\endhead
\bottomrule\noalign{}
\endlastfoot
(a) time vs.~thickness & \(N_Z\) = 2--34 cells (0.8--13 nm) & SrTiO₃
\(8\!\times\!8\!\times\!N_Z\) & \(384^2\) & 4--68 (\(2\)/cell,
\(\Delta z\,1.95\) Å) & \(8^2\) \\
(b) memory vs.~grid & \(G\) = \(128^2\)--\(1024^2\) & SrTiO₃
\(8\!\times\!8\!\times\!26\) (\(\approx\!10\) nm) & swept & 52 &
\(8^2\) \\
(c) time vs.~scan & \(P\) = \(2^2\)--\(64^2\) & SrTiO₃
\(8\!\times\!8\!\times\!26\) (\(\approx\!10\) nm) & \(128^2\) & 52 &
swept \\
\end{longtable}

\subsection{Exactness in the full-parent
limit}\label{sec-supp-exactness}

If every beam is its own parent (\(B_p=B\), \(\mathbf w=\mathbf I\)),
then \(\widehat\psi_p\) reduces to a single plane wave
\(\Psi(\mathbf h_p)\,e^{2\pi i\,\mathbf h_p\cdot(\mathbf r-\symbf{\rho})/\mathbf N}\)
and Equation~\ref{eq-part-probe} collapses term-by-term to the exact
PRISM probe Equation~\ref{eq-prism-probe}:

\begin{equation}\phantomsection\label{eq-part-exact}{
\widehat\psi_p(\mathbf r,\symbf{\rho})\,\mathcal S^{\mathrm{dt}}_p(\mathbf r)
 = \Psi(\mathbf h_p)\,e^{2\pi i\,\mathbf h_p\cdot(\mathbf r-\symbf{\rho})/\mathbf N}\,
   \mathcal S_p(\mathbf r)\,e^{-2\pi i\,\mathbf h_p\cdot\mathbf r/\mathbf N}
 = \Psi(\mathbf h_p)\,e^{-2\pi i\,\mathbf h_p\cdot\symbf{\rho}/\mathbf N}\,\mathcal S_p(\mathbf r).
}\end{equation}

Thus Section~\ref{sec-prism} is the \(B_p\!\to\!B\) limit of partitioned
PRISM and serves as its exactness oracle; partitioning is an
approximation \emph{only} through the interpolation
Equation~\ref{eq-nnw} of \(B\) beams from \(B_p\) parents.

\subsection{Validation of EELS multislice against
abTEM}\label{sec-supp-validation}

To verify the physical correctness and quantitative accuracy of the EELS
multislice implementation in \texttt{scatterem}, we performed a
cross-validation against the reference package \texttt{abtem} (version
1.0.9). We simulated a Ti-L core-loss EELS line profile across an
atomically sharp LaAlO\(_3\)/SrTiO\(_3\) interface along the {[}001{]}
growth direction. Both packages used the all-electron \texttt{gpaw} DFT
backend to calculate identical atomic radial wavefunctions, a 300 kV
beam energy, a 20 mrad probe-forming aperture, and a 20 mrad EELS
collection angle.

\begin{figure}[H]

\centering{

\includegraphics[width=0.7\linewidth,height=\textheight,keepaspectratio]{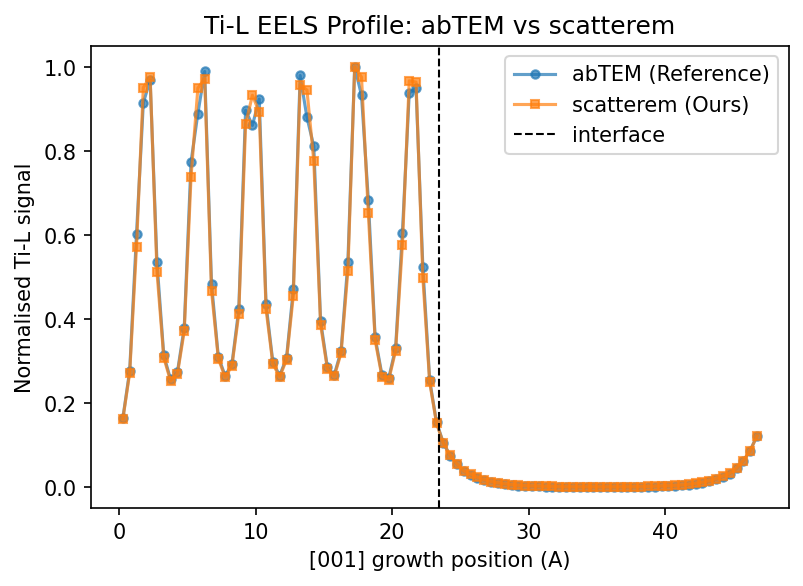}

}

\caption{\label{fig-abtem-validation}Comparison of growth-direction Ti-L
EELS profiles simulated using abTEM and scatterem across the
LaAlO\(_3\)/SrTiO\(_3\) interface.}

\end{figure}

The resulting normalized growth-direction Ti-L profiles show excellent
quantitative agreement, yielding a Pearson correlation coefficient of
\textbf{0.9986}. The comparison plot is shown in
Figure~\ref{fig-abtem-validation}. The tiny residual difference is due
to sub-pixel probe propagation differences within potential slices.

\phantomsection\label{refs}

\end{document}